# Rapid low-temperature synthesis of graphene-coated SiC substrates for remote and van der Waals epitaxy

Se H. Kim[1,†], Hanjoo Lee[1,†], Dong Gwan Kim[2,†], Donghan Kim[3,†], Seugki Kim[4], Hyunho Yang[1], Yunsu Jang[1], Jangho Yoon[1], Hyunsoo Kim[5], Seoyong Ha[5], ByoungTak Lee[6], Jung-Hee Lee[6], Roy Byung Kyu Chung[2], Hongsik Park[3*], Sungkyu Kim[4*], Tae Hoon Lee[2*], Hyun S. Kum[1*]

[1]Department of Electrical and Electronic Engineering, Yonsei University, Seoul, South Korea
[2]Department of Advanced Materials Science and Engineering, Kyungpook National University, Daegu, South Korea
[3]Department of Electronic and Electrical Engineering, Kyungpook National University, Daegu, South Korea
[4]Department of Nanotechnology and Advanced Materials Engineering, Sejong University, Seoul, South Korea
[5]LX Semicon Co., Ltd., 222, Techno 2-ro, Yuseong-gu, Daejeon, South Korea
[6]L&D Co., Ltd., 179, Daehak-ro, Yuseong-gu, Daejeon, South Korea

* Indicates corresponding authors
† These authors contributed equally: Se H. Kim, Hanjoo Lee, Dong Gwan Kim, Donghan Kim

* Corresponding author: Hongsik Park, Sungkyu Kim, Tae Hoon Lee, Hyun S. Kum
* E-mail address: hpark@ee.knu.ac.kr, sungkyu@sejong.ac.kr, thl@knu.ac.kr, hkum@yonsei.ac.kr

## Abstract

Non-conventional epitaxial techniques, such as van der Waals epitaxy (vdWE) and remote epitaxy, have attracted substantial attention in the semiconductor research community for their capability to repeatedly produce high-quality free-standing films from a single mother wafer. Successful implementation of these epitaxial techniques depends on creating a robust, uniform two-dimensional (2D) material surface. The conventional method for fabricating graphene on silicon carbide (SiC) is high-temperature graphitization. However, the extremely high temperature required for silicon sublimation (typically above 1500 °C) causes step-bunching of the SiC surface, forming non-uniform multilayer graphene stripes and an unfavorable surface morphology for epitaxial growth. Here, we developed a wafer-scale graphitization technique that allows fast synthesis of single-crystalline graphene at ultra-low temperatures by metal-assisted graphitization (MAG). We found annealing conditions that enable SiC dissociation while avoiding silicide formation, producing uniform single-crystalline graphene while maintaining the surface morphology of the substrate. The graphene thickness can be controlled by varying the metal thickness or annealing temperature, enabling remote epitaxy or vdWE. We successfully produced freestanding single-crystalline III-N (AlN, GaN) films on graphene/SiC via the 2D material-based layer transfer technique. Our results show that low-temperature graphene synthesis via MAG offers a promising route to producing large-scale ultra-wide bandgap free-standing crystalline membranes.

## 1. Introduction

Two-dimensional (2D) material-based epitaxy techniques, such as van der Waals epitaxy (vdWE) and remote epitaxy, have garnered substantial attention due to their ability to address fundamental challenges inherent in conventional heteroepitaxial techniques caused by lattice and thermal expansion mismatches[1–5]. The vdWE and remote epitaxial techniques not only alleviate these issues, but also potentially allow reuse of costly semiconductor substrates for significant cost reduction of electrical devices. These techniques enable precise exfoliation of highly functional single-crystalline membranes and allow heterogeneous integration of dissimilar materials, thereby enabling the integration of distinct electronic and photonic elements onto a single platform[1]. Numerous studies on these epitaxial techniques have been investigated for various semiconductors (Si, Ge, and III-V materials) as well as complex oxides (perovskites, spinels, and garnets)[2–4,6–8]. Successful use of graphitized silicon carbide (SiC) substrates for remote epitaxy and vdWE of GaN membranes has also been demonstrated recently[4,8]. Single-crystalline SiC is one of the most promising candidates for remote epitaxy and vdWE due to its ability to form uniform defect-free and atomically flat wafer-scale graphene through high-temperature graphitization. The graphitization process eliminates the need to transfer graphene onto the host substrate for growth, preventing process-induced damage or residues on the graphene layer that can hinder the growth of high-quality epitaxial films. However, the graphitization and subsequent formation of graphene on SiC usually occur at very high temperatures over 1500 °C[9,10]. This high-temperature requirement is a technical obstacle as the high temperature not only causes significant step-bunching of the SiC surface but also leads to non-uniform growth of graphene at the edges of the steps, generating multilayer graphene stripes and unfavorable surface morphology for epitaxial growth[9].

Recently, several groups have reported on a new graphitization method of SiC substrates utilizing a thin metal film as a catalyst to reduce the graphitization temperature. This metal-assisted graphitization (MAG) process can be achieved through the deposition of various metals on the SiC substrate, such as nickel (Ni), iron (Fe), ruthenium (Ru), cobalt (Co) and its alloys[11–25]. These studies showcase the catalytic effect of metals in breaking the Si-C covalent bond at temperatures below 1100 °C. However, many groups have observed defective interface graphene at the metal-silicide/SiC interface[15–19], suggesting that liberated carbon atoms diffuse inside the metal-silicide layer in the annealing process, and then precipitate at the silicide/SiC interface to form graphene in the cooling process. Unfortunately, the silicide formation appears to be dominant[13,16], which not only deteriorates the uniformity and roughness of graphene[13,18] but also damage the SiC substrate surface[15,17]. Notably, Lim et al. reported that SiC decomposes at approximately 450 °C using Ni as the catalyst, much below the silicide formation temperatures, resulting in the formation of 6~8 nm multilayer graphene at the metal/SiC interface due to the difference in the solubilities of carbon and silicon in Ni[19]. However, the formation mechanism and properties of the interface graphene synthesized by MAG remain unclear due to limited information on surface morphology, domain size, and graphene uniformity as previous studies have predominantly focused on graphene formed on reacted metal surfaces, silicide surfaces, and silicide/SiC interface. Thus, further investigation is needed to bridge this knowledge gap.

## 2. Results and discussion

Metal-assisted graphitization (MAG) of SiC was first reported by Juang et al.[11], in which they studied the formation of graphene by depositing a thin Ni layer on SiC and annealing at relatively low temperatures. They found that the metal film acts as a decomposition catalyst above a certain temperature (> 700 °C), which is sufficient to fully overcome the activation barrier for silicide formation. The tendency of metal to react with Si reduces the Si-C bonding energy, thereby lowering the temperature required for Si sublimation and subsequent graphitization of the SiC. We carried out a meticulous study by varying the catalyst metal element and annealing temperature as well as performing density functional theory (DFT) calculations to fully understand the MAG process. The process of MAG is illustrated in Figure 1a (see Supplementary Fig. 1 for the 4-inch wafer MAG process). A thin layer of metal is deposited onto a 4º off-axis 4H-SiC (0001) substrate using sputter deposition or e-beam evaporation for Ni, Ru, and Fe, followed by annealing in a rapid thermal annealing (RTA) chamber. An off-axis orientation for the SiC wafer was selected over the on-axis substrate to promote periodic adatom nucleation at higher-density step edges, which enhances the nucleation for epitaxial growth, and improves the crystallinity of the resulting film[4]. After cooling, any residual metal film is etched away with its respective etchants. A photograph of the resulting metal-assisted graphitized SiC wafer is shown in Figure 1b. The graphitized wafer shows a weak visible-light absorption, as indicated by the ultraviolet-visible (UV-vis) transmittance spectra (see Supplementary Fig. 2). Figure 1c illustrates the mechanism of how carbon atoms redistribute in the presence of metal catalysts at elevated temperatures, forming a graphene layer at the metal-SiC interface. The MAG process can be characterized into three distinct stages. In Stage 1, once sufficient thermal energy is introduced, the metal catalyst facilitates the decomposition of Si–C bonds. In Stage 2, as the annealing temperature increases, further dissociation occurs. The system then stabilizes as newly released C atoms either migrate outward into the metal bulk (i) or remain at the metal/SiC interface to form graphitic carbon or graphene layers (ii or iii). The redistribution of C atoms varies depending on the type of metal used due to differences in solubility and chemical affinity. Stable graphene layers are expected to form at the metal/SiC interface as a final product of the interactions (Stage 3) when appropriate metals are used.

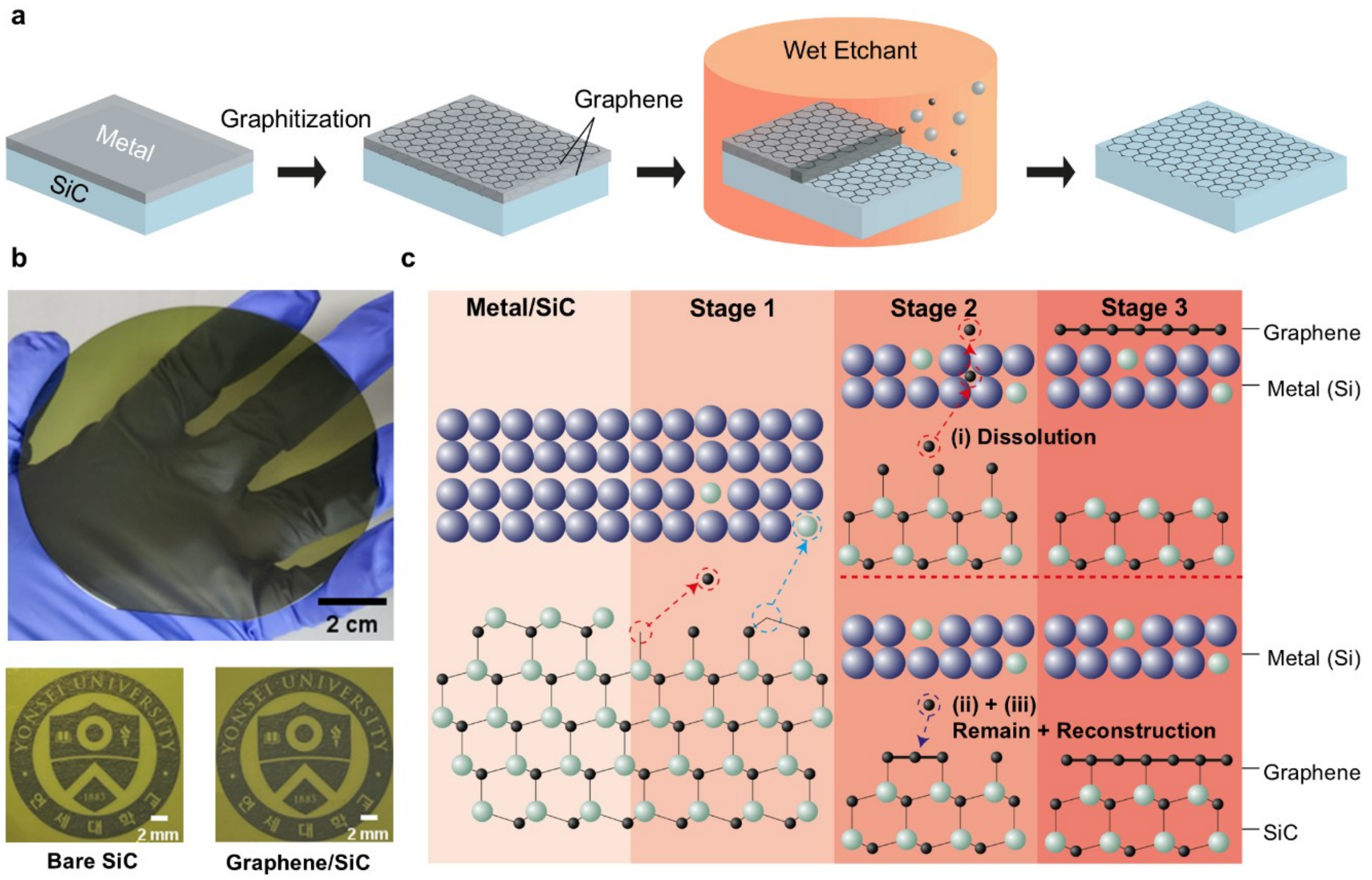

**Figure 1 | The MAG process and carbon distribution model. a,** Schematic representation of the MAG process. **b,** Photograph of graphene on a 4-inch wafer SiC (left), with zoomed-in images showing bare SiC (upper right) and multi-layer graphene on MAG-treated SiC (lower right). The darker appearance of the graphene sample indicates reduced visible light transmittance. **c,** Schematic of the carbon distribution model during the solid-state reaction in the MAG process. The presence of graphene at the metal/SiC interface depends on the metal species, which governs the redistribution of carbon toward: (i) dissolution into the metal bulk, (ii) retention at the interface to form a graphitic carbon layer, or (iii) reconstruction and recrystallization into pristine graphene. Black, light green, and dark blue spheres represent carbon (C), silicon (Si), and metal atoms, respectively.

To understand the MAG processes illustrated in Figure 1c, it is crucial to first examine the microscopic mechanism of metal-assisted graphitization of SiC. Previous studies indicate that the dissociation of SiC is driven by its reaction with metals to form thermodynamically favored metal-silicides[26–28]. Ni, Ru, and Fe, which exhibit highly negative enthalpies of silicide formation[29], enable dissociation at lower temperatures compared to the conventional graphitization process. However, the distinct properties of each metal, such as the solubility of silicon and carbon in the metal and their chemical affinity with the metal, may significantly influence not only the distribution of silicon and carbon within the metal but also the interaction of the metal with SiC or graphene. As a result, significantly different behaviors during Stages 2 and 3 are expected depending on the metal catalyst utilized, which has not been thoroughly investigated. Additionally, most previous MAG studies primarily focused on carbon distribution after the metal had converted into silicide, where precipitated carbon was observed at both the silicide/SiC interface and the silicide surface due to the solubility limitations of carbon during the cooling stage. These approaches, however, overlooked the distribution of C and Si immediately after the Si-C bond dissociation in the initial stage, where the effect of the metal is assumed to be most significant. This lack of knowledge regarding the early stage of carbon redistribution in the MAG process has resulted in outputs unsuitable for device applications due to issues such as uneven SiC consumption[16] and graphene clustering[13], both attributed to localized silicide formation[15,17,18].

To bridge the gap in understanding the microscopic processes at the early stage of the MAG process, as well as to elucidate the metal-specific differences, we employed *ab initio* molecular-dynamics (AIMD) simulations. We considered three metal elements-Ni, Ru, and Fe-as catalysts and investigated the initial stage of carbon redistribution following Si-C bond dissociation to unveil the dependence of stable graphene formation on metal types (see Supplementary Note 1 and Supplementary Fig. 3 and 4 for detailed simulation methods). As illustrated in Figure 2a, two competing pathways for carbon behavior were identified: complete dissolution (case 1), where C atoms released from the SiC substrate diffuse directly into the metal matrix, and graphitic carbon formation (case 2), where carbon atoms remain at the metal/SiC interface and form a graphene layer. As shown in Figure 2b, the comparative formation-energy analyses reveal that graphene formation at the metal/SiC interface is energetically favorable only for Si-dissolved Ni, while Fe and Ru exhibit positive formation energies, indicating that these metals prefer carbon dissolution over stabilizing a graphene layer. These findings indicate the distinctive catalytic effectiveness of Ni in promoting stable interfacial graphene formation. The structural evolution observed during AIMD simulations (Supplementary Fig. 5) supports these trends. Interesting observations include: (i) carbon and metal atoms tend to diffuse into each other to a certain extent at the metal-SiC interface, resulting in slight intermixing between the carbon and metal elements (Stage 2); (ii) carbon atoms in the graphene layer maintain their hexagonal arrangement, with the degree of structural order depending on the metal type; (iii) the van der Waals gap between the metal and graphene layer is relatively well maintained in the Ni-containing model (Stage 3), whereas Fe- or Ru-containing models show a collapse of the gap (Supplementary Fig. 6 To evaluate whether graphene can remain structurally stable under Si-free conditions during the initial stage, we compared both initial (Si-free) and Si containing models. As shown in Figure 2c and 2d, both Si-containing and pure Ni models preserve the hexagonal graphene structure, indicating effective stabilization at the metal/SiC interface during Stage 2. In contrast, Fe- and Ru-containing models exhibit pronounced graphene collapse and dissociation (see Supplementary Videos 1–3 and Supplementary Fig. 5 and 7 for quantitative analysis). These AIMD results highlight the distinctive catalytic role of the Si-dissolved Ni matrix in driving carbon reorganization and graphene-like structure formation. Ni, known for its high silicon solubility[30] and low carbon affinity[26], facilitates graphene formation at the metal/SiC interface. In contrast, graphene layers become unstable with Ru, (low Si solubility[31]) and Fe (high carbon affinity[32]), emphasizing the importance of selecting a catalytic metal with high Si solubility and low carbon chemical affinity.

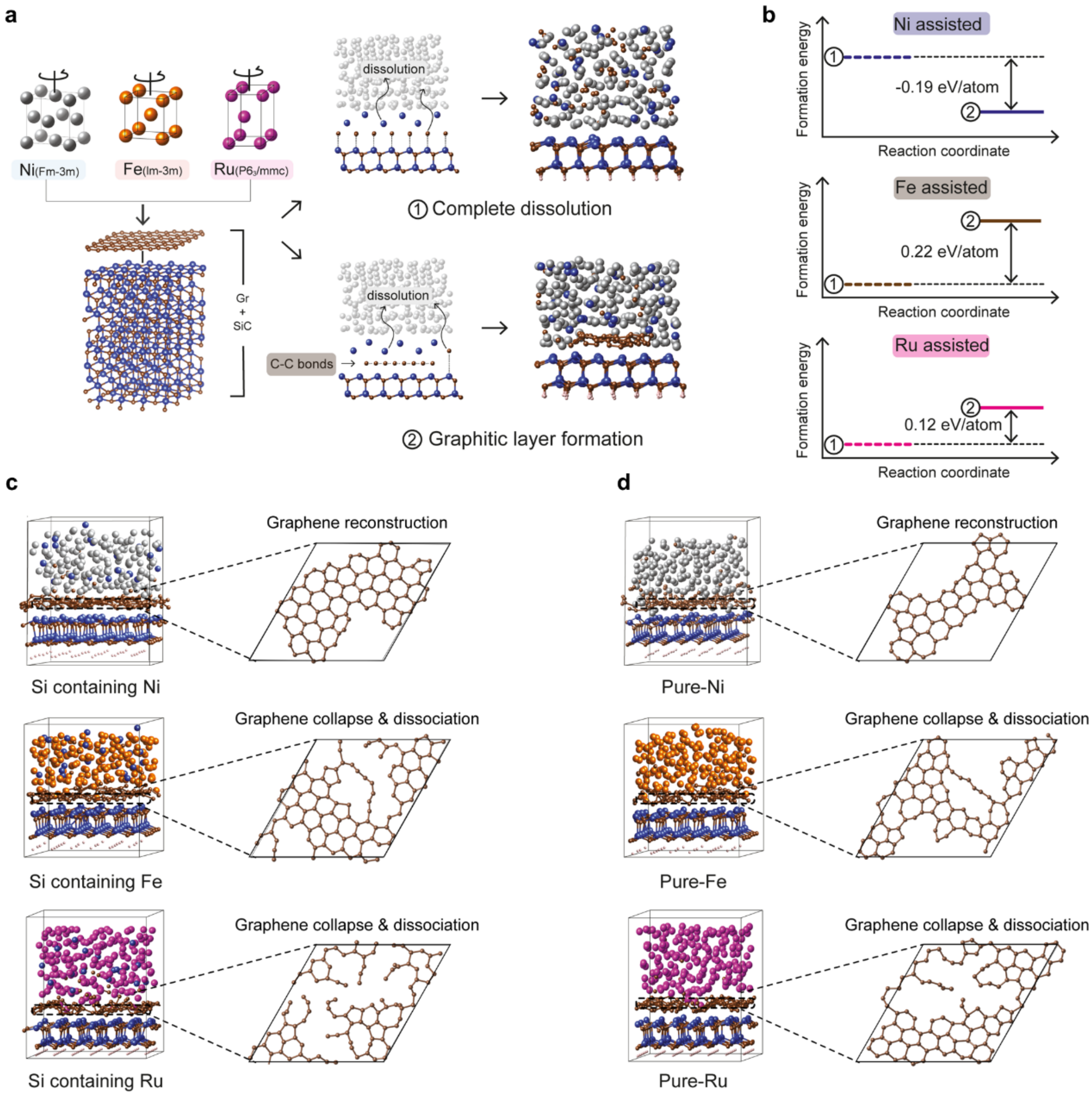

**Figure 2 | DFT modeling and atomic-scale simulation of carbon behavior at the metal-SiC interface. a,** Atomic-scale schematic of two representative pathways for carbon redistribution observed experimentally: (case 1) complete dissolution, in which carbon atoms released from the SiC substrate diffuse into the metal matrix, and (case 2) graphitic carbon formation, where carbon atoms remain at the metal/SiC interface and assemble into a graphene layer. **b,** DFT-calculated formation energies for dissolution (case 1) and graphitization (case 2) at the Ni–SiC (top), Fe–SiC (middle), and Ru–SiC (bottom) interfaces. The results highlight the distinctive catalytic role of Ni in stabilizing interfacial graphene, attributed to its high silicon solubility and low carbon affinity. **c-d,** Structural evolution of carbon at the metal–SiC interface during AIMD simulations for Si-containing (**c**) and Si-free metals (**d**). The trajectories of carbon atoms exhibit strong metal dependence: Ni promotes ordered graphene formation, while Fe and Ru induce structural disorder and collapse of the graphene layer.

These DFT results align well with our experimental results, demonstrating that only Ni enables the formation of uniform graphene at the metal/SiC interface without any silicide formation. A schematic representation of the metal's effect on MAG, highlighting the significant differences in outcomes, is illustrated in Figure 3a. To experimentally verify the catalyst effect of each metal in reducing the Si-C dissociation barrier energy during the MAG process, we performed X-ray photoelectron spectroscopy (XPS) measurements of the Si 2p and C 1s core levels after annealing at significantly low temperatures. As shown in Figure 3b, Si 2p XPS spectra confirmed the liberation of Si atoms when annealing the samples at 500 °C for Ni and Fe, and at 400 °C for Ru, showcasing a significant reduction in the energy required for Si-C bond dissociation. The formation of carbon layers on the metal surface, produced either by dissociated carbon diffusing through grain boundaries[21] or due to the low solubility of carbon in the metal matrix[20], was further confirmed by C 1s core-level XPS and Raman spectroscopy (Supplementary Fig. 8 and 9).

After confirming Si-C bond dissociation in all metal/SiC systems, we investigated the temperature-dependent phase transformations in the metals, including the formation of silicide through solid-state reactions. These transformations were characterized using Si 2p XPS and X-ray diffraction (XRD) measurements (Supplementary Fig. 10). Notably, no silicide peaks were detected for Ni annealed at 500 °C or Fe annealed at 400 °C, despite clear evidence of liberated Si atoms observed in the Si 2p spectra. However, at higher temperatures, peaks at 99.6 eV for Ni and 99.7 eV for Fe were observed, corresponding to the formation of $Ni_2Si$ and $Ni_{31}Si_{12}$ mixed phases[33], and FeSi[34], respectively. Ru, in contrast, exhibited a distinct peak at 99.8 eV, indicative of $Ru_2Si_3$ formation[35] upon annealing at 400 °C. However, the XRD analysis revealed the presence of the $Ru_2Si_3$ phase even at 350 °C, suggesting that silicide formation occurred beneath the typical XPS probing depth (~ 10 nm). Additional confirmation of structural changes was provided by metal core-level XPS, and O core-level XPS, and Raman spectroscopy (Supplementary Fig. 8 and 11). We further observed distinct effects of each metal on graphene formed at the metal surface. A distinct 2D peak in Raman spectra was only detected on Ni without the formation of a silicide layer (Supplementary Fig. 9). For Fe, graphene formation occurred only after silicide formation, where liberated carbon atoms were either contained within the metal or distributed at the metal/SiC interface. In contrast, Ru did not form a graphene layer but an amorphous carbon layer below 400 °C, despite the silicide reaction having already proceeded, potentially indicating a need for higher temperatures to crystallize the carbon layer into graphene[11,16].

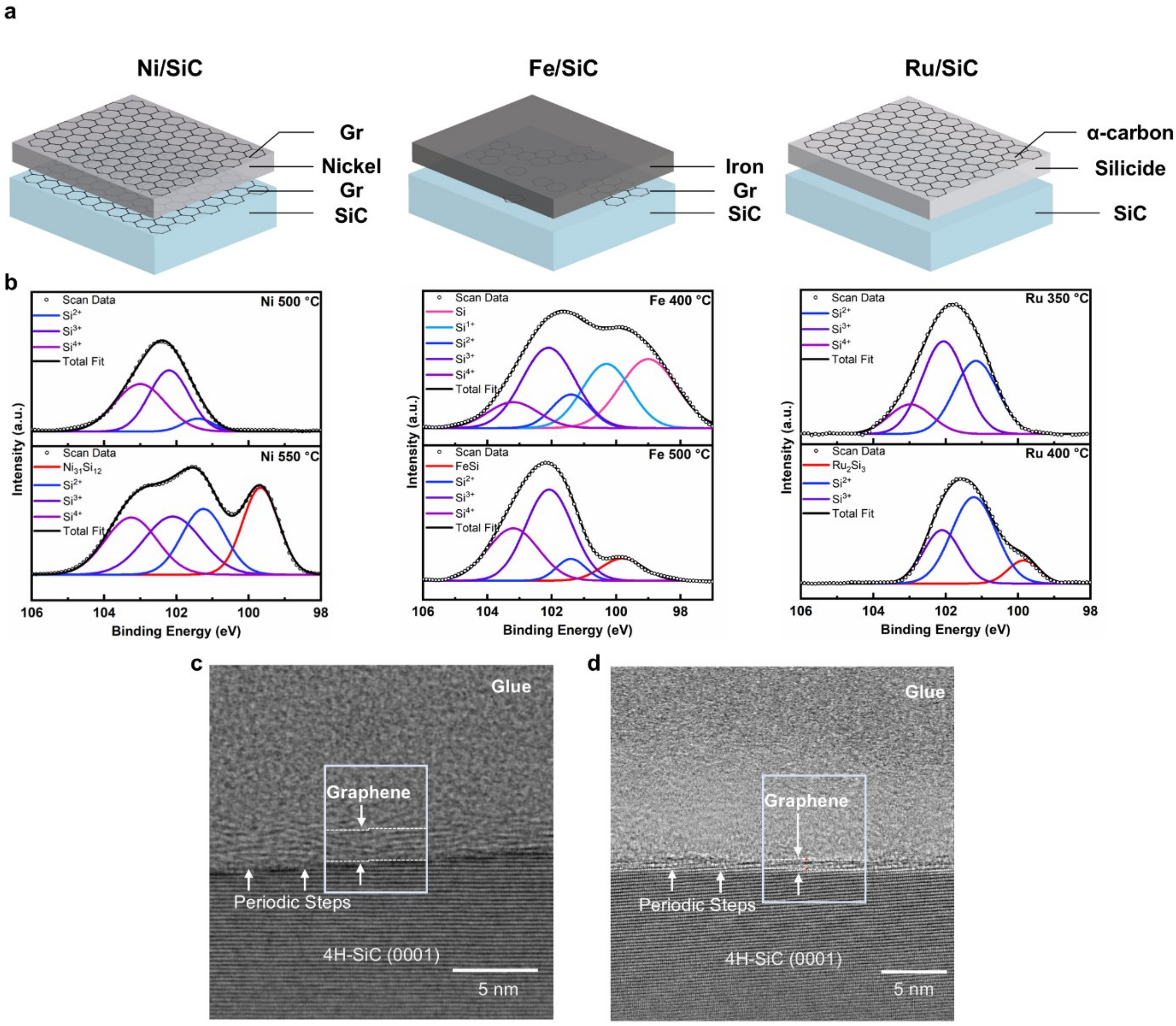

**Figure 3 | Metal effect in the MAG process, and experimental investigations. a,** Schematic representation of the metal effect in the MAG process, as calculated by DFT calculation and experimental results. The entire interface graphene layer was detected only when using Ni as a catalyst. **b,** X-ray spectra of Si 2p for various metal-SiC systems after annealing. All spectra reveal the presence of Si within the metal bulk, indicating that the Si-C bonds were decomposed by the metal catalyst. **c,** HR-TEM cross-sectional image of MAG-treated SiC, annealed at 500 °C after depositing a 50 nm Ni layer, followed by etching of metal layer. **d,** HR-TEM cross-sectional image of MAG-treated SiC, annealed at 500 °C after depositing an 8 nm Ni layer, followed by etching of the metal layer.

Since DFT calculations indicated significant differences in the possibility of graphene formation at the interface of each metal/SiC depending on the metal, further investigations of the metal/SiC interface were conducted. The residual metal film was etched away using its respective etchant after annealing at the minimum temperature needed for Si-C dissociation. Raman spectra and the optical image (OM) confirmed that a graphene layer was synthesized on SiC only when Ni was used as the catalyst, in agreement with DFT results (Supplementary Fig. 12). In contrast, no carbon layer was detected for Ru annealed at 350 °C, while an uneven amorphous carbon layer was observed for Fe annealed at 400 °C, suggesting that the carbon at the metal/SiC interface either diffused outward into the reacted region or was absorbed into the metal layer, again consistent with the DFT results. These results unambiguously verify that the choice of the metal catalyst plays a crucial role in the MAG process. Among the three metals studied, only Ni was able to form wafer-scale graphene at the metal/SiC interface, without any silicide or carbide formation.

After selecting the appropriate metal (Ni) based on the above results, further experimental studies were conducted to validate the carbon distribution model at both low temperatures (stage 1 in Figure 1c) and relatively high temperatures (stage 2 in Figure 1c). We further lowered the annealing temperature not only to control the thickness of the graphene but also to explore the relationship between annealing temperature and Si-C bond dissociation. As shown in Figure 3c and Supplementary Fig. 13, transmission electron microscopy (TEM) analysis confirmed the formation of multi-layer graphene (MLG, more than three monolayers thick) at the Ni/SiC interface after annealing at 500 °C for ~1 second, while few-layer graphene (FLG, between one to two monolayers thick) formed at 320 °C for ~1 second. These results are consistent with the UV-vis transmittance spectra obtained at the macroscopic scale after etching with a respective etchant (Supplementary Fig. 14). Furthermore, when the annealing temperature was increased to 550 °C, we observed not only inhomogeneous silicide reactions but also clustering of graphene, as confirmed by atomic force microscopy (AFM), scanning electron microscopy (SEM), and Raman spectroscopy (Supplementary Fig. 15). These findings align with observations reported in the literatures[15,20].

Furthermore, we reduced the thickness of Ni from 50 nm to 8 nm to investigate the relationship between metal thickness and graphene thickness. As dissociated Si atoms are homogeneously distributed within the Ni lattice[36], we hypothesized that reducing the metal thickness would decrease the reactivity between SiC and Ni. This reduction is attributed to the lower gradient-driven diffusion of Si atoms at thinner metal layers, resulting from shorter diffusion pathways or a diminished driving force for Si-C dissociations, thereby leading to thinner graphene on the SiC surface. As shown in Figure 3d, the thickness of graphene on the SiC surface decreased to approximately single-layer. Once again, the UV-vis transmittance confirmed decreasing transmittance due to the increased graphene thickness as the metal thickness increased (see Supplementary Fig. 16). Overall, the results indicate that the MAG process is primarily temperature-driven, but the thickness of the graphene layer can be controlled by adjusting the metal layer thickness or the annealing temperature.

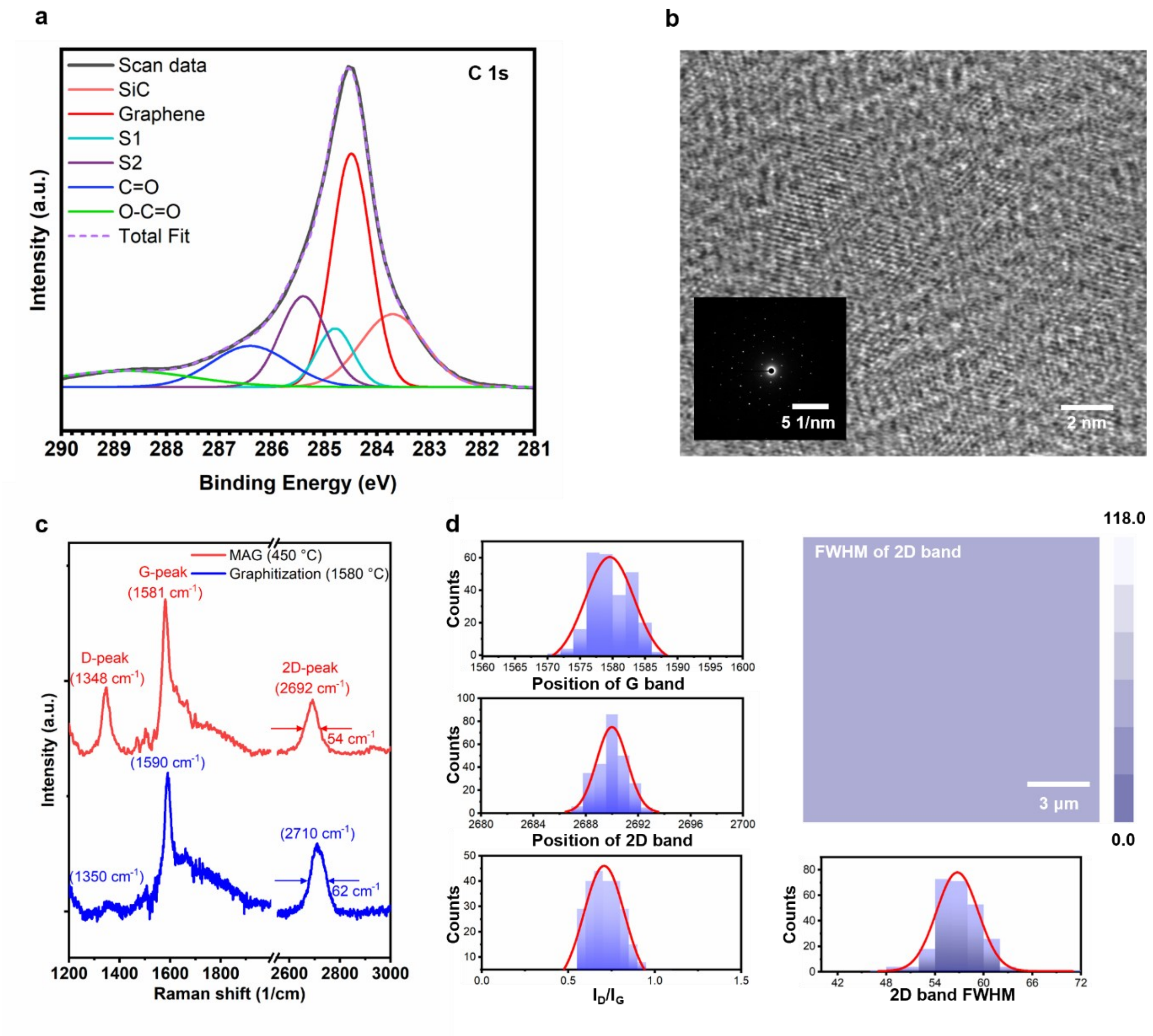

**Figure 4 | Wafer-scale single-crystalline graphene formed on SiC via the MAG process. a,** XPS spectra of the C 1s region, showing graphene on the entire SiC. **b,** Plan-view TEM image with SAED patterns of graphene transferred onto TEM grid, verifying the single-crystalline nature of the graphene. **c,** Raman spectra displaying clear D, G, and 2D peaks, confirming the presence of graphene on SiC, with quality comparable to that of graphene synthesized via high-temperature graphitization. **d,** Raman maps of the G and 2D peak positions, $I_D/I_G$ ratio, and 2D FWHM, demonstrating the uniformity of graphene quality and thickness. A total of 256 points were measured within a 15 μm × 15 μm Raman mapping area.

Next, we characterized the MLG formed on SiC by etching away the Ni layer graphitized at 500 °C for ~1 second. The presence of graphene was confirmed by XPS, Raman spectroscopy, and plan-view TEM. As shown in Figure 4a, the C 1s spectrum consists of a SiC bulk component at 283.7 eV, the $sp^2$ C=C bond of graphene at 284.5 eV, and two surface buffer layer components, S1 (284.8 eV) and S2 (285.4 eV). The S1 and S2 peaks exhibited an intensity ratio of approximately 1:2, with full widths at half maximum of (FWHM) values of 0.8 eV and 1.0 eV, respectively. These peaks originate from the graphene buffer layer; S1 corresponds to carbon atoms partially bonded to underlying Si atoms, while S2 is attributed to $sp^2$ bonded carbon exhibiting stronger in-plane bonding within the buffer layer[37,38]. Additionally, the XPS Ni 2p spectra and energy-dispersive X-ray spectroscopy (EDS) confirmed the complete removal of Ni, as it was not detected on the surface (Supplementary Fig. 17). The graphene was also characterized using plan-view TEM (See Experimental details). The TEM image and selected area electron diffraction (SAED) pattern reveals a well-aligned honeycomb lattice, providing clear evidence of single-crystalline dominant graphene, as shown in Figure 4b.

To evaluate the characteristics of graphene and compare them with the high-temperature graphitized sample on a macroscopic scale, Raman spectroscopy was performed. As shown in Figure 4c, both spectra confirmed the formation of graphene, as evidenced by the characteristic D (~1350 $cm^{-1}$), G (~1580 $cm^{-1}$), and 2D (~2700 $cm^{-1}$) bands[39]. The MAG-treated graphitized sample exhibited a more pronounced D peak ($I_D/I_G \approx 0.5$) compared to the high-temperature graphitized sample ($I_D/I_G \approx 0.1$), indicating a higher density of structural imperfections such as point defects, surface corrugations, or substrate-induced perturbations[40]. Despite the increased D-band intensity, the clear SAED pattern confirmed the preservation of single-crystalline order, suggesting that these defects are not associated with grain boundaries or extended structural disorder. Instead, the enhanced D band likely originates from atomically localized defects, such as vacancies or point imperfections[41,42], as well as structural or electronic perturbations induced by the underlying substrate[40]. Both samples exhibited distinct G and 2D bands, commonly used as indicators of crystalline quality and thickness of graphene[39–45]. The positions of the G and 2D bands of the MAG-treated graphene remained nearly unchanged compared to pristine graphene, indicating negligible doping and strain effects[40,43]. These results suggest that MAG-treated SiC effectively preserves the structural quality of graphene without inducing significant lattice perturbations. In contrast, the high temperature graphitized sample exhibited a marked upshift in both the G-band (~10 $cm^{-1}$) and the 2D band (~30 $cm^{-1}$), indicating considerable lattice distortion driven by charge transfer or residual strain. The FWHM of the 2D band serves as a quantitative measure not only of graphene thickness, ranging from single-layer graphene (typically, 27.5 ± 3.8 $cm^{-1}$) to four-layer graphene (typically, 63.1 ± 1.6 $cm^{-1}$)[9,46], but also of the long-range crystalline order[42,47]. The narrower 2D-band FWHM observed in the MAG-treated graphene relative to the high-temperature counterpart suggests similar crystalline quality at comparable thicknesses. As shown in Figure 4d, Raman mapping revealed highly uniform G and 2D positions (1580 ± 0.52 $cm^{-1}$ and 2690 ± 0.18 $cm^{-1}$, respectively). The 2D band also exhibited a consistent FWHM (averaging 57 ± 0.18 $cm^{-1}$) across the substrate. This spatial uniformity provides strong evidence for the high structural homogeneity and consistent thickness of the graphene film across the entire SiC wafer. Furthermore, complete graphene coverage across the 4-inch SiC substrate was confirmed (see Raman spectra and camera image in Supplementary Fig. 18), demonstrating the capability to produce 4-inch graphene directly on a semiconductor substrate using a very low-temperature, ultra-fast process, that relies solely on RTA and sputtering, which are techniques fully compatible with Si CMOS industry standards.

The results of MAG suggest that the SiC substrate can be used as an ideal substrate for producing single-crystalline free-standing high-quality III-nitride films by remote epitaxy and vdWE. Remote epitaxy is an advanced epitaxial technique where epitaxial growth occurs on a substrate covered with a 2D material[1]. The partial transparency of the 2D layer to Coulombic interactions allows adatoms to electrostatically interact with the underlying substrate[1–3]. At the substrate-epitaxial membrane interface, the 2D material and its van der Waals gap effectively eliminate dislocations or cracks in the epitaxial membrane caused by lattice strain relaxation, which has been a persistent challenge in conventional techniques for achieving high-quality epitaxial devices[2,48,49]. In contrast, vdWE involves epitaxial growth on 2D materials. These slippery interfaces enable strain relaxation, allowing the growth of materials with significant lattice mismatches greater than 60%[1]. Furthermore, in both 2D-based epitaxial techniques, epitaxial membranes are freestanding on the substrate and only bound to the substrate via a weak van der Waals force. This characteristic enables the fabrication of freestanding membranes through 2D material-assisted layer transfer (2DLT)[7]. These approaches are particularly advantageous, as they allow the repeated reuse of expensive substrates while producing multiple single-crystalline membranes. A schematic of the detailed 2D-based epitaxy and 2DLT process is shown in Supplementary Fig. 19.

To demonstrate the applicability of our MAG-treated SiC substrate, III-nitrides films were epitaxially grown. III-nitride materials such as GaN and AlN are promising candidates for power and RF devices as well as light emitting diodes due to their exceptional intrinsic material properties[50,51]. These include wide direct bandgaps (3.4 eV for GaN and 6.2 eV for AlN)[51,52], high breakdown electric field (4.9 MV $cm^{-1}$ for GaN and 15.4 MV $cm^{-1}$ for AlN)[47], and high electron mobility. These materials are particularly well-suited for 2D-based epitaxial growth on graphitized SiC, as their hexagonal lattice arrangement aligns well with SiC compared to other available substrates, allowing high quality single-crystalline growth via remote epitaxy. Furthermore, the graphitization of SiC produces nearly pristine graphene, whose preserved hexagonal lattice structure enhances the growth of c-plane III-nitrides films, offering significant benefits for vdWE[49]. Finally, the direct synthesis of graphene on SiC eliminates the need for wet-transferred graphene synthesized on metal foils via CVD, which often introduces defects such as wrinkles, holes, interfacial contamination, and organic residues. These defects can disrupt the remote interaction between the substrate and the remote epitaxial film, as well as between the graphene and the van der Waals epitaxial film[2].

Although the advantages of the graphitized SiC template for 2D-based epitaxy are clear, high-quality III-N membrane growth on graphene faces challenges due to its low chemical reactivity[53]. The high surface migration rate of group III metals on slippery graphene prevents nuclei from stabilizing, which results in highly defective epitaxial films due to insufficient nucleation sites[49]. High-quality single-crystalline AlN film growth via quasi-vdWE on graphitized SiC has been reported, wherein plasma treatment was applied to enhance nucleation by introducing defects on the graphene surface[49,54,55]. Although partial exfoliation of AlN layers synthesized through quasi-vdWE has been demonstrated[54], full-area transfer remains challenging due to the strengthened interfacial bonding between the AlN epilayer and the defect-modified graphene. As a result, previous studies have primarily focused on synthesizing crack-free AlN layers that leverage the stress relaxation benefits of graphene. Since the membrane exfoliation yield improves with a uniform graphene layer covering the entire surface on the substrate, untreated graphene is required to successfully produce freestanding membranes. The successful exfoliation of GaN on non-defect-induced graphitized SiC via 2D-assisted epitaxy was first reported in 2014, achieved not only by engineering the epitaxial growth strategy on a 2D surface but also by utilizing the periodical step edges of SiC[4]. These step edges, with terrace widths typically ranging from 550 to 800 nm and step heights from 30 to 45 nm, remain after the step bunching induced by high-temperature processes[56]. These periodic step edges generate uniform fluctuations in electric potential, providing energetically favorable nucleation sites for adatoms and enabling the growth of single-crystalline GaN[4]. In this regard, our MAG-graphitized approach is expected to offer significant advantages. The MAG process is conducted at low temperatures, which avoids step bunching and preserves the naturally periodic nanoscale terraces of 4° offcut 4H-SiC (see AFM images and line profile in Supplementary Fig. 20). These terraces exhibit widths below 7.2 nm and step heights under 0.5 nm[57]. To verify this advantage, we grew AlN on both high-temperature graphitized and MAG SiC substrates. As illustrated in Figure 5a, growth on high-temperature graphitized SiC resulted in very rough non-uniform polycrystalline AlN growth following the step-bunched surface morphology, whereas the MAG SiC substrate resulted in smooth single-crystalline AlN growth.

High-quality AlN films were grown on both MAG-treated MLG/SiC substrate and FLG/SiC substrate by vdWE and remote epitaxy, respectively. For both samples, electron backscatter diffraction (EBSD) maps with SEM images and XRD scans confirmed a (002) wurtzite orientation across a large area, as shown in Figure 5b-5d (see in-plane crystallinity in Supplementary Fig. 21). These results indicate that single-crystalline AlN can be grown via remote epitaxial seeding from FLG/SiC and van der Waals epitaxial seeding from MLG/SiC (see the illustrated image in Supplementary Fig. 22). The FWHM of AlN (002) peaks in XRD scans broadened with increasing graphene thickness on SiC, suggesting lower seeding efficiency compared to remote epitaxy. Prior studies reported on the successful exfoliation of high-quality AlN on graphitized SiC grown via 2D-based epitaxy, achieving an AlN (002) FWHM of 3600 arcsec[53], and partial exfoliation of high-quality AlN via quasi-vdWE, with a (002) FWHM of 792 arcsec and (102) FWHM of 1044 arcsec[54]. In contrast, our study demonstrated significantly improved AlN quality on MAG-treated SiC, with remote epitaxy achieving a (002) FWHM of 759 arcsec and a (102) FWHM of 833 arcsec, while vdWE-grown AlN exhibited broader peaks, with a (002) FWHM of 1744 arcsec and a (102) FWHM of 2136 arcsec, at a film thickness below 300 nm (see Supplementary Fig. 21). The exfoliated AlN layer exhibiting the lowest (002) FWHM is presented in Figure 5e, highlighting the high crystalline quality achieved through our approach. We concluded that the dramatically enhanced crystallinity of AlN can be attributed to the presence of tightly packed and periodically stepped SiC. These features provide highly energetically favorable nucleation sites, which are effective for both remote epitaxy[4] and vdWE[49,55].

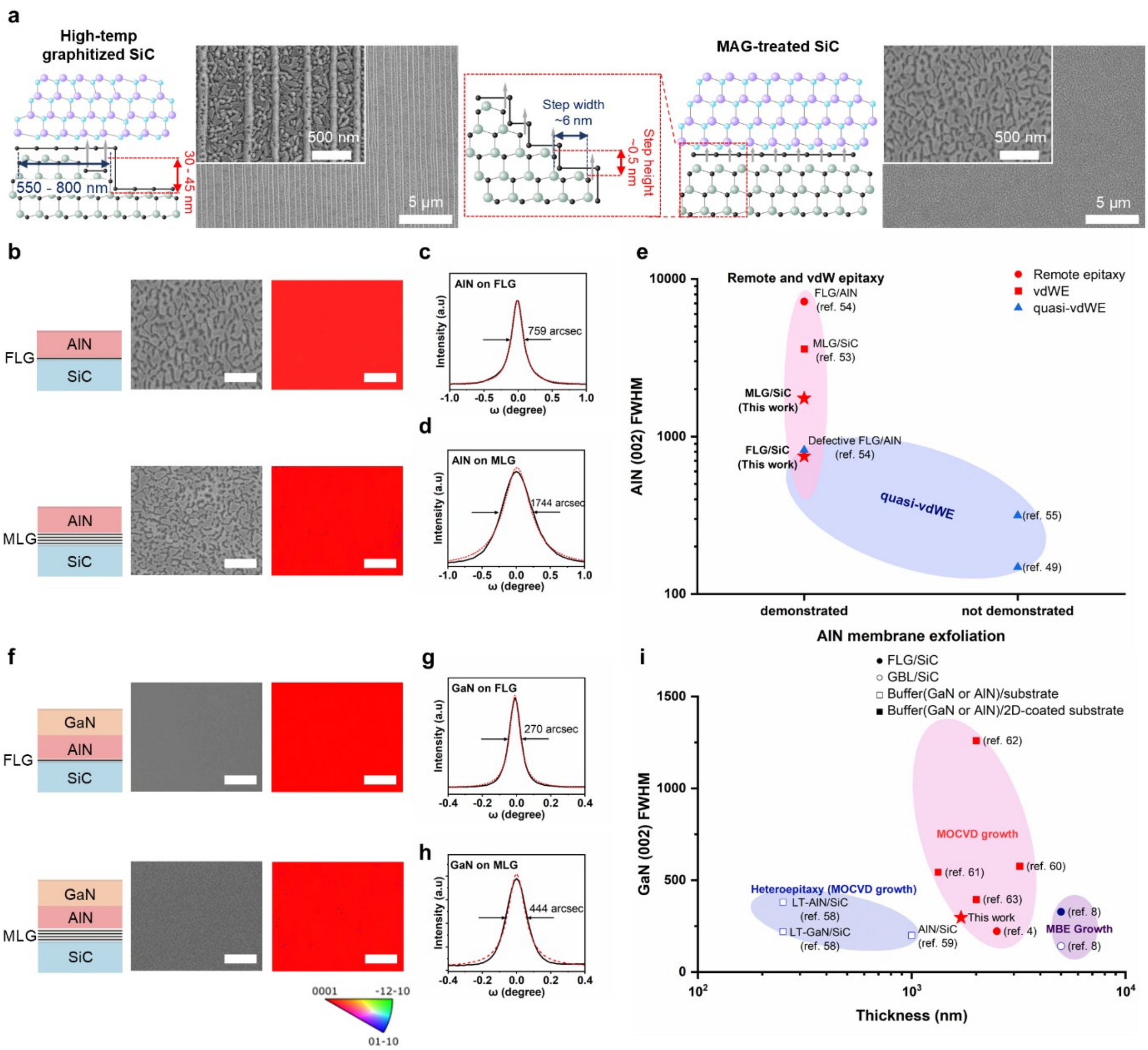

**Figure 5 | Free-standing III-nitride membranes synthesized on MAG-treated graphene/SiC templates. a,** Schematic illustration and SEM images of 2D-assisted epitaxy on MAG-treated graphitized SiC and high-temperature graphitized SiC. Unlike high-temperature graphitization, MAG-treated SiC avoids step bunching, offering potential advantages for growing high-quality membranes through 2D-assisted epitaxy. **b,** Schematic illustration of the grown film structure (left), SEM image (center), and EBSD map (right), showing the highly single-crystalline nature of the AlN membrane over a large scale on FLG/SiC and MLG/SiC templates, respectively. **c-d,** XRD ω-scan of both templates. The FWHM of the AlN (002) peak increases with graphene layer thickness. **e**, XRD benchmark of the exfoliated AlN membrane, showing the lowest (002) FWHM, as referenced in the main text. **f**, Schematic illustration of the grown film structure (left), SEM image (center), and EBSD map (right), showing the highly single-crystalline nature of the GaN membrane over a large scale on the AlN/FLG/SiC template. **g-h,** XRD ω-scans of GaN membranes grown on AlN/FLG/SiC and AlN/MLG/SiC templates, respectively, showing no significant differences. **i,** XRD FWHM benchmark of the exfoliated GaN membrane, confirming the high crystalline quality achieved. The scale bars in all SEM images and EBSD map are 1 μm.

After confirming the successful growth of a single-crystalline AlN layer on MAG-treated SiC, we used it as a buffer layer to grow single-crystalline GaN films. In conventional epitaxy, AlN layers (a-axis: 3.112 Å) are commonly employed as intermediate layers to address the lattice mismatch between GaN (a-axis: 3.189 Å) and SiC (a-axis: 3.073 Å), thereby enhancing the quality of GaN. While graphene as a buffer layer effectively relaxes the lattice strain of GaN thin films, the crystallinity of GaN grown on AlN synthesized via 2D-assisted epitaxy still requires thorough investigation. To evaluate this, we grew GaN on each 2D-assisted epitaxial AlN templates and conducted XRD and EBSD measurements to determine the crystallinity on a macroscopic scale. The EBSD maps with SEM images and XRD scans verified the (002) wurtzite orientation over a large area, indicating single crystallinity of the grown GaN film on both AlN templates, as shown in Figure 5f-h (see in-plane crystallinity in Supplementary Fig 24). Notably, the FWHM of GaN (002) were 270 and 444 arcsec for remote epitaxy and vdWE, respectively, while the GaN (102) FWHM were 509 and 760 arcsec for remote epitaxy and vdWE, respectively. These results are comparable not only to those of AlN-buffer-assisted GaN films on conventional SiC via MOCVD[58,59], but also to the remote epitaxial growth of GaN on graphitized SiC and other 2D-coated substrates[4,8,60–63], as shown in Figure 5i. The crystal quality can be further improved by optimizing the growth recipe. After confirming the single crystallinity of the GaN layer, we utilized 2DLT to exfoliate both samples. Following the deposition of the adhesion layer (Ti) and stressor layer (Ni) on the surface of the GaN samples, mechanical exfoliation was carried out using thermal release tape (TRT) as a handling layer. The strain energy generated by the Ni stressor guided crack propagation precisely along the AlN/graphene interface, facilitated by the weak van der Waals bonds between the GaN/AlN and the graphitized SiC. As a result, freestanding GaN/AlN membranes were successfully obtained for both MAG-treated templates (see Supplementary Fig. 25).

**4. Conclusions**

In conclusion, we demonstrate low-temperature ultrafast growth of wafer-scale graphene on a 4-inch SiC substrate below 500 °C, with precise control on the number of graphene layers by changing the metal thickness and annealing temperature. The resulting graphene grown by MAG exhibits overall uniform characteristics comparable to graphene synthesized by traditional high-temperature graphitization. Moreover, the III-nitride films grown on MAG-treated SiC via 2D-assisted epitaxy demonstrate quality comparable to those grown on high-temperature graphitized SiC and conventional substrates such as sapphire and SiC by MOCVD. We found that the metal employed significantly influences the presence of graphene layers on SiC, with Ni being the only catalyst capable of synthesizing uniform graphene on SiC. These results offer valuable insights into the general mechanisms of the MAG process, providing a more accessible approach for synthesizing wafer-scale graphene which can be readily used for remote epitaxy or vdWE, as well as produce stripe-free single-crystalline graphene at wafer-scale. Our approach not only dramatically reduces the barrier of preparing epitaxial graphene coated single-crystalline substrates, but also significantly enhances the crystallinity of the freestanding single-crystalline membranes for heterogeneous integration.

## Experimental details

### Computational details

The first-principles calculations were performed using the projected augmented wave (PAW) plane-wave basis, implemented in the Vienna ab initio simulation package (VASP)[64]. An energy cutoff of 520 eV was employed and the atomic positions were optimized using the conjugate gradient scheme without any symmetric restrictions, until the maximum force on each of them was less than 0.01 eV/Å[65]. All atoms were relaxed to their equilibrium positions when the change in energy on each atom between successive steps converged to $1\times10^{-6}$ eV/-atom. The heterostructure was modeled with an 8×8×1 grid for k-point sampling. The generalized gradient approximation (GGA) exchange-correlation (XC) DFT functional Perdew-Burke-Ernzerhof (PBE) was employed for geometrical optimization and electronic structure calculations[66]. The slab models had dangling bonds on the vacuum surface terminated by pseudo-hydrogen atoms with appropriate fractional charges to avoid surface states. To determine the vacuum level, dipole corrections are introduced to compensate for the artificial dipole moment at the open ends (20 Å vacuum space along the c-axis) arising from the periodical boundary condition imposed in these calculations[67]. Ab initio molecular dynamics (AIMD) simulations were performed in supercells using DFT calculations with a gamma-centered k-point. The time step was set to 3 fs. Simulations for 3 ps were run with a time step of 3 fs to study the dynamic graphitization process. The temperature of the simulation system was controlled at 2273 K using the Nosé–Hoover thermostat[68,69].

### Sample preparation

The single-crystalline 4° offcut 4H–SiC (0001) substrates were supplied by Cree, Inc.. To remove organic contaminants, substrates were cleaned sequentially for 5 min in acetone, 5 min Iso Propyl Alcohol in an ultrasound bath, and finally dried by nitrogen gun.

**Metal deposition and graphene formation/transfer method**

After preparing the SiC substrate, we deposited Ni, Fe, and Ru onto each SiC substrate to investigate the effect of metal catalysts on the interface graphene layer. Ni was deposited on the substrate using a DC magnetron sputtering apparatus with an Ar plasma at room temperature (JURA deposition apparatus made by Vakuum Servis Ltd., Czech Republic). The sputtering was carried out in an Ar atmosphere (pressure 0.5 mTorr) with DC power 100 W. Resulting deposition rate was 2.5 nm/min. Fe and Ru were deposited on the substrate using an e-beam evaporator (Korea Vacuum Tech., Korea Republic). The base pressure was maintained below $5 \times 10^{-7}$ torr, with a deposition rate of 0.6 Å/s.

Following metal deposition, the samples underwent rapid thermal annealing (RTA). The base pressure during RTA was maintained at $7 \times 10^{-3}$ torr, with 1 torr of $N_2$ added to prevent metal oxidation. The samples were annealed for 3 minutes to determine the temperature required to decompose the Si-C bond and assess whether it induces a phase change of metals into silicide. X-ray photoelectron spectroscopy (XPS; K-alpha, Thermo Scientific Inc.) employing an aluminum anode (spot size: 400 μm; energy resolution: FWHM $\leq$ 0.5 eV for Ag $3d_{5/2}$; pass energy: 40 eV; step size: 0.1 eV) was employed to analyze the elemental diffusion and chemical bonding at the surface. In parallel, X-ray diffraction (XRD; Rigaku SmartLab, Cu $k_{\alpha 1}$, $\lambda = 1.54051$ Å) was used to identify crystalline phases and monitor structural transformations within the metal/SiC structure. Raman spectroscopy (Horiba Jobin Yvon, LabRam Aramis) equipped with a 532 nm wavelength laser was utilized to analyze phase transformations of metals and to confirm the presence of graphene. Following the reaction, the interfacial graphene layer was analyzed using high-resolution transmission electron microscopy (HRTEM, JEOL ARM200F). To further investigate the interface characteristics, the samples were immersed in a 40% w/v Ferric Chloride ($FeCl_3$) solution (for Ni, and Fe) for 5 minutes or a 5% w/v Sodium Hypochlorite (NaOCl) solution (for Ru) for 5 minutes. The samples were then characterized using optical microscopy (OM), scanning electron microscopy (SEM, JEOL, JSM-IT-500HR), Ultraviolet-visible-near-infrared spectrophotometer (UV-vis-NIR, JASCO, V-650), atomic force microscopy (AFM, Park Systems, NX-10), TEM, and Raman spectroscopy. Raman analysis was performed after normalizing all spectra and subtracting the signal from bare SiC (specifically the SiC band at 1520 $cm^{-1}$, which exhibits the maximum intensity across all samples within the spectral range of 1200 $cm^{-1}$ to 3000 $cm^{-1}$) to enable precise analysis of the graphene signal.

After graphene synthesis, polyvinyl alcohol (PVA) was drop-casted onto the surface and baked at 80 °C for 5 minutes to form an adhesion layer. A TRT was then attached, facilitating the detachment of graphene through mechanical exfoliation. The exfoliated graphene was transferred onto TEM grid. The TRT was removed by baking the sample at 130 °C, and the graphene was revealed by dipping the sample into deionized water.

**SiC high-temperature graphitization**

The wafer was loaded into an Aixtron VP508 reactor for graphitization. It was first cleaned in a hydrogen environment for 30 minutes at 1520 °C, followed by annealing at 1580 °C in a 700 Torr argon ambient for 10 minutes.

**III-N epitaxial Growth and exfoliation**

The GaN/AlN hetero-structure was epitaxially grown on a graphene/SiC substrate by a metal-organic chemical vapor deposition (MOCVD) equipped with a vertical showerhead-type chamber from Sysnex Co., Ltd. The MOCVD reactor maintained a stable pressure of 30 Torr with hydrogen as a carrier gas throughout the growth process. Initially, the AlN buffer layer was grown at 1,050 °C for 40 mins on FLG/SiC and MLG/SiC. Subsequently, GaN layer was grown by a single-step growth at 1,100 °C for 10 mins on each AlN layer. In this growth process, trimethylgallium, trimethylaluminum, and ammonia were used as the sources of gallium, aluminum, and nitrogen, respectively. The resulting GaN/AlN epilayers were exfoliated using a Ti/Ni stressor stack with a handling layer. A 50 nm thick Ti layer was deposited as an adhesion layer for the Ni stressor layer via e-beam evaporation, followed by the deposition of a 3.5 μm thick Ni stressor layer using DC magnetron sputtering under an argon ambient. After the deposition of Ti/Ni layers, thermal release tape (TRT) was attached as a handling layer. Finally, the TRT/Ti/Ni/epi stack was lifted from the edges, enabling precise and controlled exfoliation of the GaN/AlN epilayers.

**III-N membrane characterizations**

The plan-view, cross-sectional, and EBSD images of the grown and exfoliated samples were obtained using a field-emission SEM system (SU8220, Hitachi). X-ray diffraction characterization was carried out using an XRD system (Empyrean, Malvern Panalytical, Cu $K_{\alpha 1}$, $\lambda$ = 1.54051 Å) to identify crystalline phases and assess structural evolution.

### Credit authorship contribution statement

**Se H. Kim:** Conceptualization, Writing – original draft, Visualization, Methodology, Investigation. **Hanjoo Lee:** Investigation, Writing – original draft. **Dong Gwan Kim:** DFT simulation, Writing – original draft. **Donghan Kim:** Investigation, Visualization, Validation. **Seugki Kim:** Visualization, Validation. **Hyunho Yang:** Visualization, Validation. **Yunsu Jang:** Visualization, Validation. **Jangho Yoon:** Validation. **Hyunsoo Kim:** Resources. **Seoyong Ha**: Resources. **ByuongTak Lee**: Resources. **Jung-Hee Lee**: Resources. **Roy Byung Kyu Chung**: Resources. **Hongsik Park:** Validation, Resources. **Sungkyu Kim:** TEM, Validation. **Taehoon Lee:** DFT, Validation, Resources. **Hyun S. Kum:** Conceptualization, Methodology, Investigation, Resources, Writing – review & editing, Project administration, Supervision.

### Declaration of Competing Interest

The authors declare that they have no known competing financial interests or personal relationships that could have appeared to influence the work reported in this paper.

### Data availability

Data will be made available on request.

### Acknowledgments

The team at Yonsei University would like to acknowledge support from the National Research Foundation of Korea (NRF) (grant no. RS-2023-00222070, grant no. RS-2024-00445081, grant no. RS-2024-00451173), Samsung Electronics, and LX Semicon.

### References

[1] H. Kum, D. Lee, W. Kong, H. Kim, Y. Park, Y. Kim, Y. Baek, S. H. Bae, K. Lee, J. Kim, *Nat. Electron.* **2019**, *2*, 439–450.

[2] H. Kim, C. S. Chang, S. Lee, J. Jiang, J. Jeong, M. Park, Y. Meng, J. Ji, Y. Kwon, X. Sun, W. Kong, H. S. Kum, S. H. Bae, K. Lee, Y. J. Hong, J. Shi, J. Kim, *Nat. Rev. Methods Primers* **2022**, *2*, 40.

[3] C. S. Chang, K. S. Kim, B.-I. Park, J. Choi, H. Kim, J. Jeong, M. Barone, N. Parker, S. Lee, X. Zhang, K. Lu, J. M. Suh, J. Kim, D. Lee, N. M. Han, M. Moon, Y. S. Lee, D.-H. Kim, D. G. Schlom, Y. J. Hong, J. Kim, *Sci. Adv.* **2023**, *9*, eadj5379

[4] J. Kim, C. Bayram, H. Park, C. W. Cheng, C. Dimitrakopoulos, J. A. Ott, K. B. Reuter, S. W. Bedell, D. K. Sadana, *Nat. Commun.* **2014**, *5*, 4836.

[5] J. Narayan, *Acta Mater.* **2013**, *61*, 2703–2724.

[6] H. S. Kum, H. Lee, S. Kim, S. Lindemann, W. Kong, K. Qiao, P. Chen, J. Irwin, J. H. Lee, S. Xie, S. Subramanian, J. Shim, S. H. Bae, C. Choi, L. Ranno, S. Seo, S. Lee, J. Bauer, H. Li, K. Lee, J. A. Robinson, C. A. Ross, D. G. Schlom, M. S. Rzchowski, C. B. Eom, J. Kim, *Nature* **2020**, *578*, 75–81.

[7] Y. Kim, S. S. Cruz, K. Lee, B. O. Alawode, C. Choi, Y. Song, J. M. Johnson, C. Heidelberger, W. Kong, S. Choi, K. Qiao, I. Almansouri, E. A. Fitzgerald, J. Kong, A. M. Kolpak, J. Hwang, J. Kim, *Nature* **2017**, *544*, 340–343.

[8] K. Qiao, Y. Liu, C. Kim, R. J. Molnar, T. Osadchy, W. Li, X. Sun, H. Li, R. L. Myers-Ward, D. Lee, S. Subramanian, H. Kim, K. Lu, J. A. Robinson, W. Kong, J. Kim, *Nano Lett.* **2021**, *21*, 4013–4020.

[9] K. V. Emtsev, A. Bostwick, K. Horn, J. Jobst, G. L. Kellogg, L. Ley, J. L. McChesney, T. Ohta, S. A. Reshanov, J. Röhrl, E. Rotenberg, A. K. Schmid, D. Waldmann, H. B. Weber, T. Seyller, *Nat. Mater.* **2009**, *8*, 203–207.

[10] C. Berger, Z. Song, X. Li, X. Wu, N. Brown, C. Naud, D. Mayou, T. Li, J. Hass, A. N. Marchenkov, E. H. Conrad, P. N. First, W. A. de Heer, *Science* **2006**, *312*, 1191–1196.

[11] Z. Y. Juang, C. Y. Wu, C. W. Lo, W. Y. Chen, C. F. Huang, J. C. Hwang, F. R. Chen, K. C. Leou, C. H. Tsai, *Carbon* **2009**, *47*, 2026–2031.

[12] W. Yuan, C. Li, D. Li, J. Yang, X. Zeng, *J. Nanomater.* **2011**, *2011*, 319624.

[13] F. Iacopi, N. Mishra, B. V. Cunning, D. Goding, S. Dimitrijev, R. Brock, R. H. Dauskardt, B. Wood, J. Boeckl, *J. Mater. Res.* **2015**, *30*, 609–616.

[14] H. I. Røst, R. K. Chellappan, F. S. Strand, A. Grubisic-Cabo, B. P. Reed, M. J. Prieto, L. C. Tanase, L. De Souza Caldas, T. Wongpinij, C. Euaruksakul, T. Schmidt, A. Tadich, B. C. C. Cowie, Z. Li, S. P. Cooil, J. W. Wells, *J. Phys. Chem. C* **2021**, *125*, 4243–4252.

[15] E. Escobedo-Cousin, K. Vassilevski, I. Nikitina, N. Wright, A. O'Neill, A. Horsfall, J. Goss, in *Mater. Sci. Forum*, Trans Tech Publications Ltd, **2012**, *717–720*, 629–632.

[16] P. Macháč, T. Fidler, S. Cichoň, L. Mišková, *Thin Solid Films* **2012**, *520*, 5215–5218.

[17] P. Macháč, T. Fidler, S. Cichoň, V. Jurka, *J. Mater. Sci.: Mater. Electron.* **2013**, *24*, 3793–3799.

[18] E. Escobedo-Cousin, K. Vassilevski, T. Hopf, N. Wright, A. O'Neill, A. Horsfall, J. Goss, in *Mater. Sci. Forum*, Trans Tech Publications Ltd, **2014**, *778–780*, 1162–1165.

[19] S. Lim, J. S. Oh, Y. Kwon, B. S. An, J. H. Bae, T. H. Kim, M. H. Park, H. S. Kim, C. W. Yang, *J. Nanosci. Nanotechnol.* **2016**, *16*, 10853–10857.

[20] Y. Kwon, B. S. An, C. W. Yang, *Mater. Charact.* **2018**, *140*, 259–264.

[21] A. Hähnel, V. Ischenko, J. Woltersdorf, *Mater. Chem. Phys.* **2008**, *110*, 303–310.

[22] J. Hofrichter, B. N. Szafranek, M. Otto, T. J. Echtermeyer, M. Baus, A. Majerus, V. Geringer, M. Ramsteiner, H. Kurz, *Nano Lett.* **2010**, *10*, 36–42.

[23] A. Delamoreanu, C. Rabot, C. Vallee, A. Zenasni, *Carbon* **2014**, *66*, 48–56.

[24] S. P. Cooil, F. Song, G. T. Williams, O. R. Roberts, D. P. Langstaff, B. Jørgensen, K. Høydalsvik, D. W. Breiby, E. Wahlström, D. A. Evans, J. W. Wells, *Carbon* **2012**, *50*, 5099–5105.

[25] P. Macháč, T. Hrebicek, *J. Mater. Sci.: Mater. Electron.* **2017**, *28*, 12425–12431.

[26] T. C. Chou, A. Joshi, J. Wadsworth, *J. Mater. Res.* **1991**, *6*, 796–803.

[27] W. M. Tang, Z. X. Zheng, H. F. Ding, Z. H. Jin, *Mater. Chem. Phys.* **2002**, *74*(3), 258–264

[28] M. Wu, H. Huang, Y. Wu, X. Wu, *Ceram. Int.* **2024**, *50*, 17930–17939.

[29] M. E. Schlesinger, *Chem. Rev.* **1990**, *90*, 607–628.

[30] B. P. Nash, A. Nash, *J. Phase Equilib.* **1987**, *8*, 6

[31] L. Perring, F. Bussy, J. C. Gachon, P. Feschotte, *J. Alloys Compd.* **1999**, *284*, 198–205.

[32] C. Mattevi, H. Kim, M. Chhowalla, *J. Mater. Chem.* **2011**, *21*, 3324–3334.

[33] Y. Cao, L. Nyborg, U. Jelvestam, *Surf. Interface Anal.* **2009**, *41*, 471–483.

[34] N. Ohtsu, M. Oku, K. Satoh, K. Wagatsuma, *Appl. Surf. Sci.* **2013**, *264*, 219–224.

[35] S. van Vliet, A. Troglia, E. Olsson, R. Bliem, *Appl. Surf. Sci.* **2023**, *608*, 155139, DOI 10.1016/j.apsusc.2022.155139.

[36] Y. Hoshino, S. Matsumoto, T. Nakada, Y. Kido, *Surf. Sci.* **2004**, *556*, 78–86

[37] K. V. Emtsev, F. Speck, T. Seyller, L. Ley, J. D. Riley, *Phys. Rev. B* **2008**, *77*, 155303, DOI 10.1103/PhysRevB.77.155303.

[38] N. Zebardastan, J. Bradford, J. Lipton-Duffin, J. MacLeod, K. Ostrikov, M. Tomellini, N. Motta, *Nanotechnology* **2023**, *34*, 105601.

[39] J. Röhrl, M. Hundhausen, K. V. Emtsev, T. Seyller, R. Graupner, L. Ley, *Appl. Phys. Lett.* **2008**, *92*, 201918.

[40] Z. H. Ni, W. Chen, X. F. Fan, J. L. Kuo, T. Yu, A. T. S. Wee, Z. X. Shen, *Phys. Rev. B* **2008**, *77*, 115416.

[41] A. Eckmann, A. Felten, A. Mishchenko, L. Britnell, R. Krupke, K. S. Novoselov, C. Casiraghi, *Nano Lett.* **2012**, *12*, 3925–3930.

[42] A. Zandiatashbar, G. H. Lee, S. J. An, S. Lee, N. Mathew, M. Terrones, T. Hayashi, C. R. Picu, J. Hone, N. Koratkar, *Nat. Commun.* **2014**, *5*, 3186.

[43] T. M. G. Mohiuddin, A. Lombardo, R. R. Nair, A. Bonetti, G. Savini, R. Jalil, N. Bonini, D. M. Basko, C. Galiotis, N. Marzari, K. S. Novoselov, A. K. Geim, A. C. Ferrari, *Phys. Rev. B* **2009**, *79*, 205433.

[44] E. B. Barros, N. S. Demir, A. G. Souza Filho, J. Mendes Filho, A. Jorio, G. Dresselhaus, M. S. Dresselhaus, *Phys. Rev. B* **2005**, *71*, 165422.

[45] J. A. Robinson, C. P. Puls, N. E. Staley, J. P. Stitt, M. A. Fanton, K. V. Emtsev, T. Seyller, Y. Liu, *Nano Lett.* **2009**, *9*, 964–968.

[46] Y. Hao, Y. Wang, L. Wang, Z. Ni, Z. Wang, R. Wang, C. K. Koo, Z. Shen, J. T. L. Thong, *Small* **2010**, *6*, 195–200.

[47] A. C. Ferrari, D. M. Basko, *Nat. Nanotechnol.* **2013**, *8*, 235–246.

[48] S. H. Bae, K. Lu, Y. Han, S. Kim, K. Qiao, C. Choi, Y. Nie, H. Kim, H. S. Kum, P. Chen, W. Kong, B. S. Kang, C. Kim, J. Lee, Y. Baek, J. Shim, J. Park, M. Joo, D. A. Muller, K. Lee, J. Kim, *Nat. Nanotechnol.* **2020**, *15*, 272–276.

[49] Y. Wang, S. Yang, H. Chang, W. Li, X. Chen, R. Hou, J. Yan, X. Yi, J. Wang, T. Wei, *Appl. Surf. Sci.* **2020**, *520*, 146358.

[50] D. K. Pradhan, D. C. Moore, A. M. Francis, J. Kupernik, W. J. Kennedy, N. R. Glavin, R. H. Olsson, D. Jariwala, *Nat. Rev. Mater.* **2024**, *9*, 731–732.

[51] C. Zhou, A. Ghods, V. G. Saravade, P. V. Patel, K. L. Yunghans, C. Ferguson, Y. Feng, B. Kucukgok, N. Lu, I. T. Ferguson, *ECS J. Solid State Sci. Technol.* **2017**, *6*, Q149.

[52] J. Gong, J. Zhou, P. Wang, T. H. Kim, K. Lu, S. Min, R. Singh, M. Sheikhi, H. N. Abbasi, D. Vincent, D. Wang, N. Campbell, T. Grotjohn, M. Rzchowski, J. Kim, E. T. Yu, Z. Mi, Z. Ma, *Adv. Electron. Mater.* **2023**, *9*, 2201309.

[53] Y. Xu, B. Cao, Z. Li, D. Cai, Y. Zhang, G. Ren, J. Wang, L. Shi, C. Wang, K. Xu, *ACS Appl. Mater. Interfaces* **2017**, *9*, 44001.

[54] F. Liu, T. Wang, X. Gao, H. Yang, Z. Zhang, Y. Guo, Y. Yuan, Z. Huang, J. Tang, B. Sheng, Z. Chen, K. Liu, B. Shen, X.-Z. Li, H. Peng, X. Wang, *Sci. Adv.* **2023**, *9*, eadf8484.

[55] Z. Chen, Z. Liu, T. Wei, S. Yang, Z. Dou, Y. Wang, H. Ci, H. Chang, Y. Qi, J. Yan, J. Wang, Y. Zhang, P. Gao, J. Li, Z. Liu, *Adv. Mater.* **2019**, *31*, 1807345.

[56] P. Avouris, C. Dimitrakopoulos, *Mater. Today* **2012**, *15*, 86–97.

[57] W. Chen, M. A. Capano, *J. Appl. Phys.* **2005**, *98*, 114907.

[58] C. Bayram, J. L. Pau, R. McClintock, M. Razeghi, *J. Appl. Phys.* **2008**, *104*, 083512.

[59] Z. J. Reitmeier, S. Einfeldt, R. F. Davis, X. Zhang, X. Fang, S. Mahajan, *Acta Mater* **2010**, *58*, 2165.

[60] S. Lee, J. Kim, B. I. Park, H. I. Kim, C. Lim, E. Lee, J. Y. Yang, J. Choi, Y. J. Hong, C. S. Chang, H. S. Kum, J. Kim, K. Lee, H. Kim, G. C. Yi, *Appl. Phys. Lett.* **2024**, *125*, 252102.

[61] Y. Kobayashi, K. Kumakura, T. Akasaka, T. Makimoto, *Nature* **2012**, *484*, 223–227.

[62] Y. Yu, T. Wang, X. Chen, L. Zhang, Y. Wang, Y. Niu, J. Yu, H. Ma, X. Li, F. Liu, G. Deng, Z. Shi, B. Zhang, X. Wang, Y. Zhang, *Light Sci. Appl.* **2021**, *10*, 117.

[63] N. R. Glavin, K. D. Chabak, E. R. Heller, E. A. Moore, T. A. Prusnick, B. Maruyama, D. E. Walker, D. L. Dorsey, Q. Paduano, M. Snure, *Adv. Mater.* **2017**, *29*, 1701838.

[64] G. Kresse, J. Furthmüller, *Phys. Rev. B* **1996**, *54*, 11169.

[65] P. E. Blöchl, *Phys. Rev. B* **1994**, *50*, 17953.

[66] J. P. Perdew, K. Burke, M. Ernzerhof, *Phys. Rev. Lett.* **1996**, *77*, 3865.

[67] G. Makov, M. C. Payne, *Phys. Rev. B* **1995**, *51*, 4014

[68] S. Nosé, *J. Chem. Phys.* **1984**, *81*, 511.

[69] W. G. Hoover, *Phys. Rev. A* **1985**, *31*, 1695.

# SUPPLEMENTARY INFORMATION

## Rapid low-temperature synthesis of graphene-coated SiC substrates for remote and van der Waals epitaxy

Se H. Kim, Hanjoo Lee, Dong Gwan Kim, Donghan Kim, Seugki Kim, Hyunho Yang, Yunsu Jang, Jangho Yoon, Hyunsoo Kim, Seoyong Ha, ByoungTak Lee, Jung-Hee Lee, Roy Byung Kyu Chung, Hongsik Park, Sungkyu Kim, Tae Hoon Lee, Hyun S. Kum

**Supplementary Note 1 | DFT Simulation modeling workflow**

The slab models exhibited dangling bonds on the vacuum-exposed surface, which were passivated using pseudo-hydrogen atoms with appropriate fractional charges to prevent the formation of surface states. To determine the vacuum level, dipole corrections were applied to compensate for the artificial dipole moment generated at the open ends (with a 20 Å vacuum space along the c-axis), arising from the periodical boundary conditions imposed in these calculations.

To identify an ideal hetero-interface, various combinations of substrate and film Miller indices were investigated[1]. Based on criteria such as mismatch strain, supercell area, and lattice vector length, domain-matched heterostructures between SiC and graphene with lattice mismatches below 4% were selected (Supplementary Fig. 3c). A similar matching procedure was used to construct interfaces between the SiC/graphene and the metal layers.

To emulate the graphene-formaton phenomenon during the metal-assisted graphitization (MAG) process, AIMD simulations were performed in three main steps, as illustrated in Supplementary Fig. 4: (i) a Si-containing metal layer was generated by substituting 10 at.% of the metal atoms with Si; (ii) the SiC/Graphene/Si-dissolved metal heterostructure was relaxed through energy minimization to stabilize the initial atomic configuration of each model; and (iii) random displacements were introduced to the carbon atoms in the graphene layer to study the stabilization effects of metal catalysts arising from interface interactions beetween the metal and graphene layers[2–4].

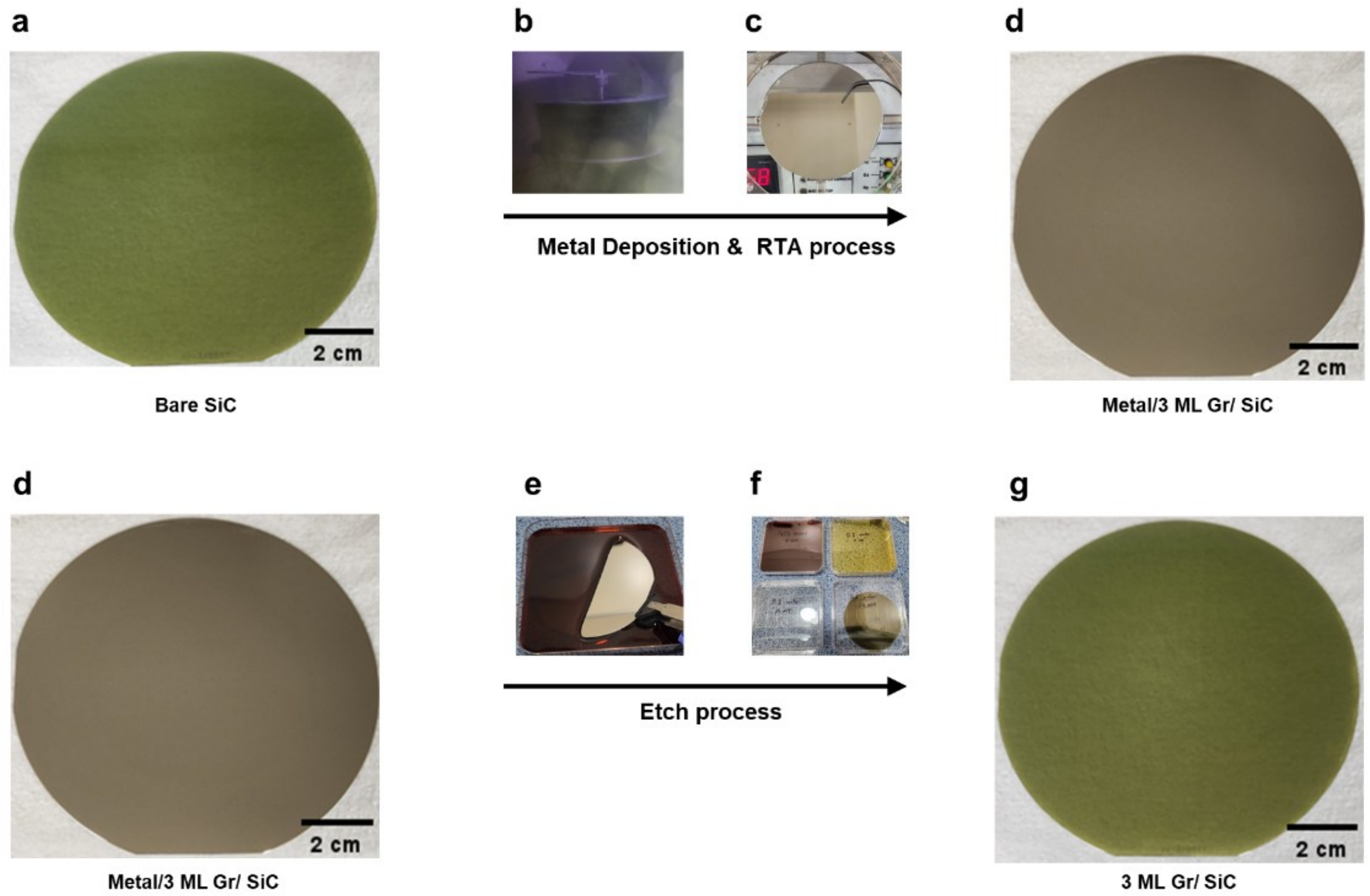

**Supplementary Fig. 1 | Detailed MAG process. a,** The SiC wafer is cleaned sequentially in acetone (5 minutes), and isopropyl alcohol (IPA) for 5 minutes each using an ultrasound bath and then dried with a nitrogen gun. **b,** Ni is deposited onto the SiC by sputtering at a pressure of $5 \times 10^{-4}$ Torr in an Ar ambient (50 sccm) for 20 minutes. **c,** The sample is annealed in a rapid thermal annealing (RTA) chamber under specific conditions: a heating speed of 12 °C/s, annealing for 3 minutes, and cooling by turning off the power. **d,** Graphene forms at the Ni/SiC interface through the MAG process. **e,** Ni is etched away using ferric chloride ($FeCl_3$) to expose the interface graphene. **f,** Once all visible traces of Ni are removed, the sample is gently agitated in fresh $FeCl_3$, followed by rinsing in deionized water. The surface is kept wet to prevent redeposition of Ni residues. **g,** The final result of the MAG process demonstrates graphene formation on the entire 4-inch SiC substrate.

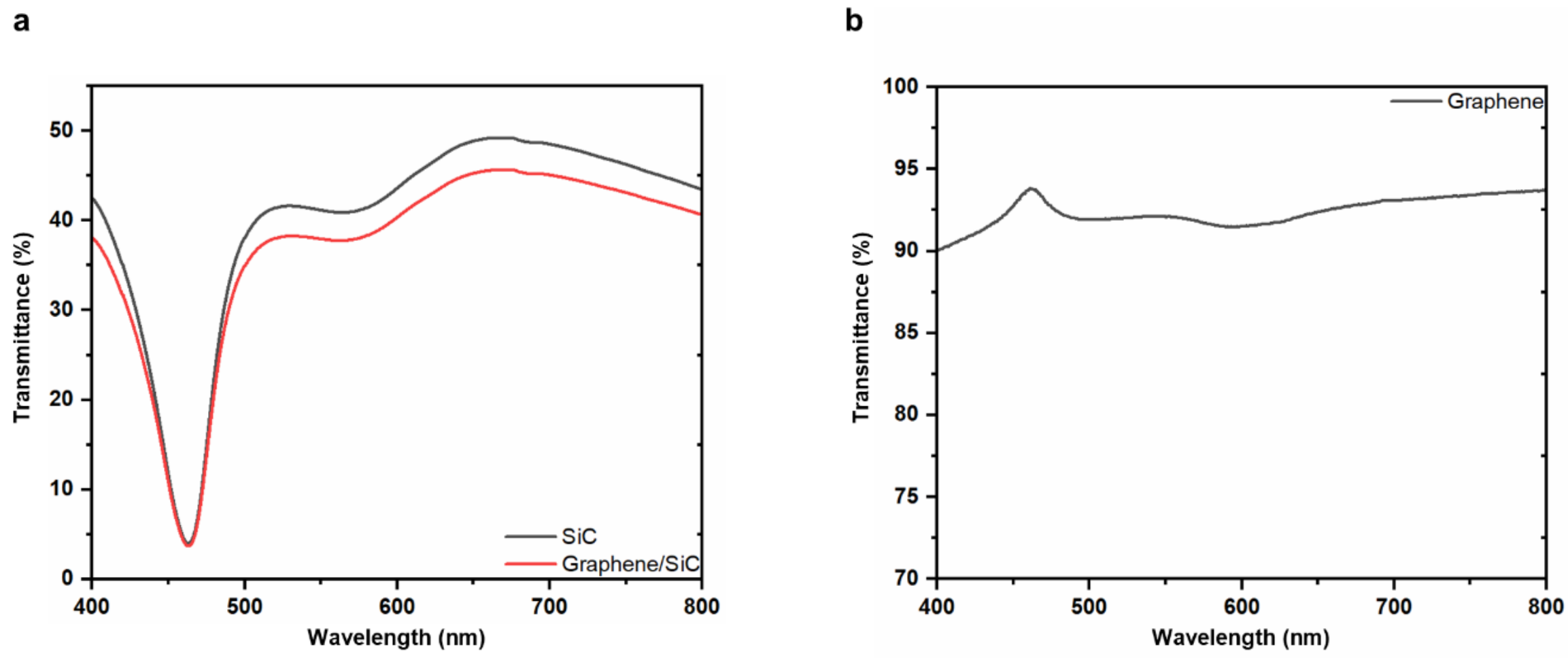

**Supplementary Fig. 2 | UV-vis transmittance spectra in the wavelength range of 400 nm - 800 nm. a,** UV-vis transmittance spectra of a SiC without graphene (red) and with graphene (black), showing reduced transmittance due to the graphene layer. **b,** The transmittance of graphene in the visible light is measured at 92.1% at 550 nm, indicating the presence of approximately 3.4 graphene monolayers on the SiC.

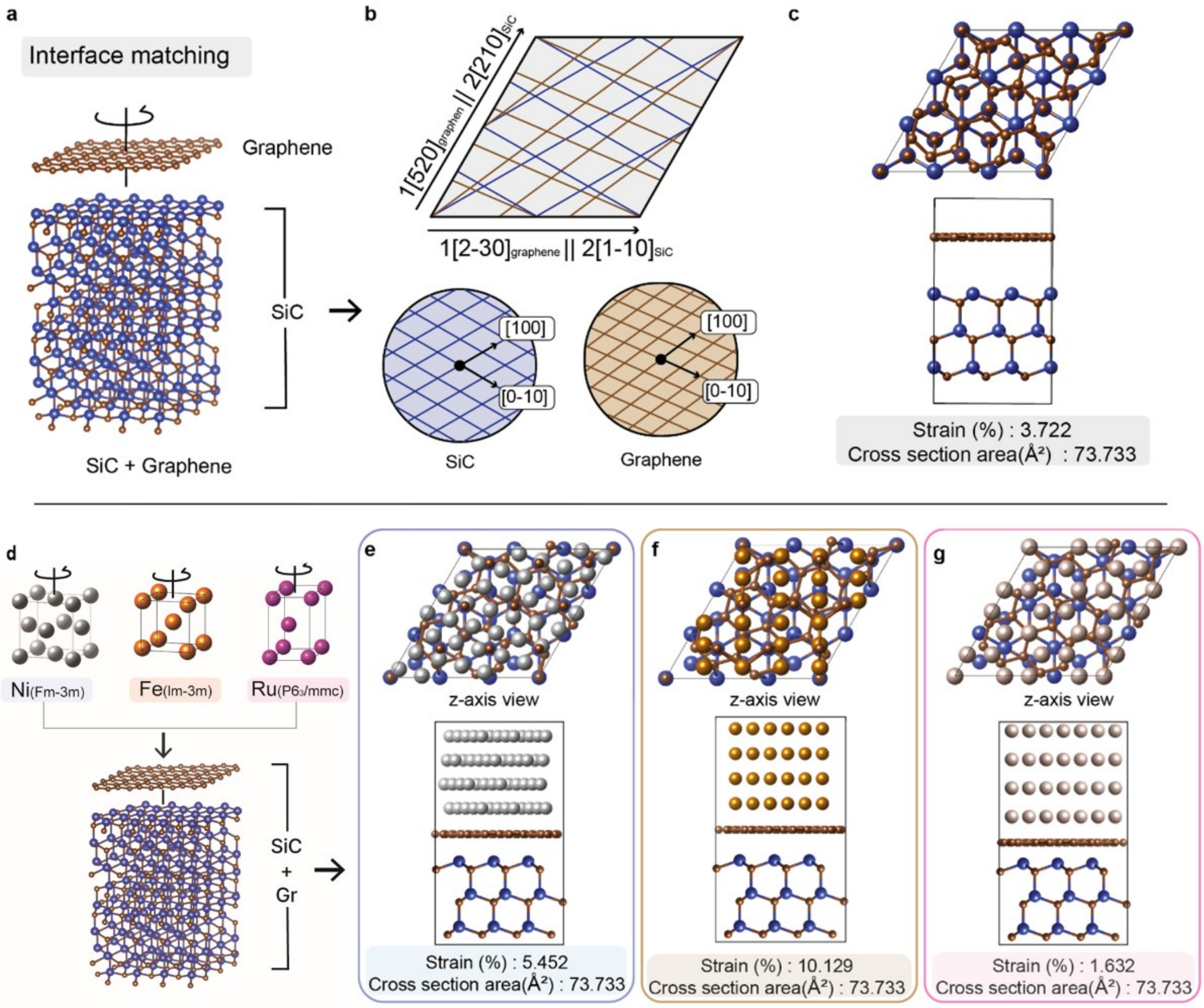

**Supplementary Fig. 3 | Interface matching for modeling the metal/graphene/SiC structures. a,** Schematic illustration of interface construction for the graphene/SiC structure. **b,** Domain-matched supercell construction for the SiC (blue) and graphene (brown) layers, with the corresponding lattice unit vectors. **c,** Top and side views of the optimized interface configuration (Si atoms shown in blue; C atoms in dark brown). **d,** Schematic illustration of the interface-matching process for the metal/graphene/SiC system, along with the crystal structures of Ni (Fm-3m), Fe (Im-3m), and Ru ($P6_3$/mmc) used in the modelling. **e-g,** Domain-matched heterostructures employed in the AIMD simulations: Ni (5.452% strain), Fe (10.129% strain), and Ru (1.632% strain), each with a cross-sectional area of 43.256 Å$^2$.

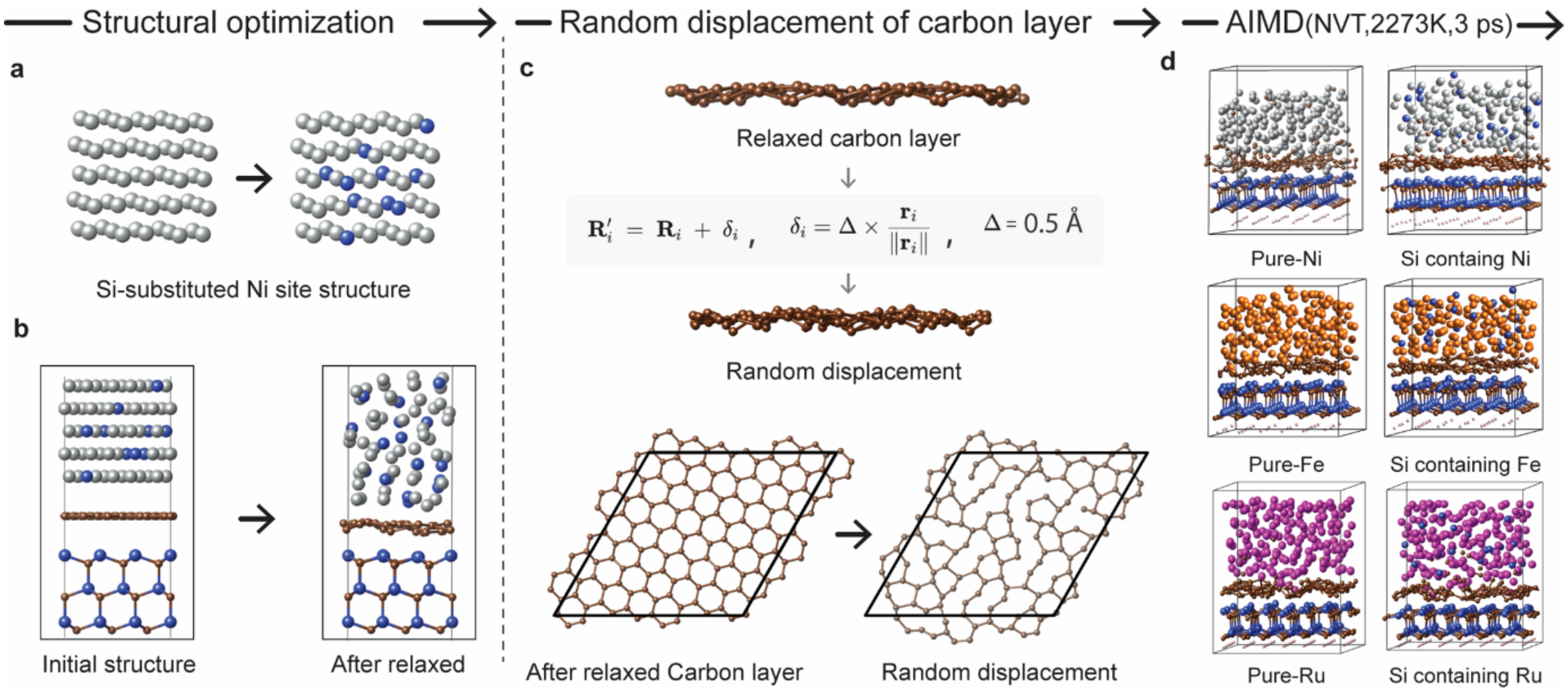

**Supplementary Fig. 4 | Modeling workflow and AIMD simulations for the metal/graphene/SiC heterostructures. a,** Formation of Si-containing metal layers. **b,** Example of a relaxed configuration of the metal/graphene/SiC heterostructure model after energy minimization. **c,** Introduction of randomly-displaced carbon atoms in the graphene layer to study the stabilization effects of metal catalysts. **d,** Model configurations used in this study. AIMD simulations were performed for both pure and Si-containing metal layers for each metal.

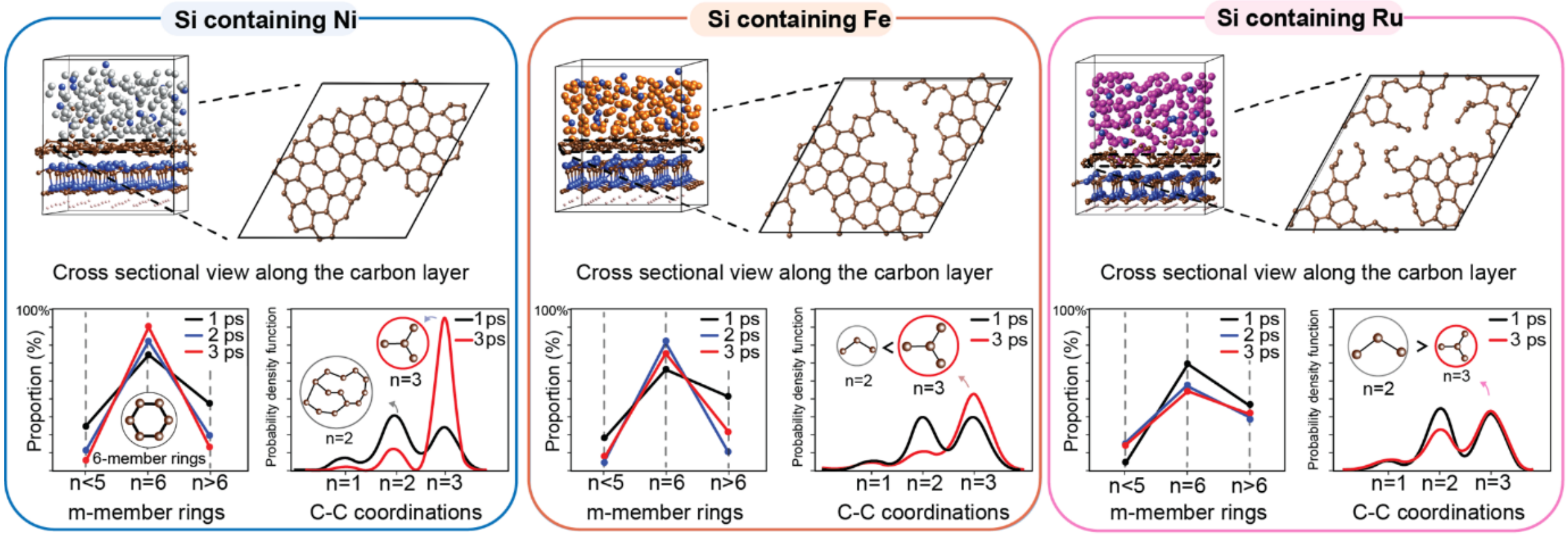

**Supplementary Fig. 5 | AIMD simulations of Si-alloyed metal layers.** Structural evolution of a carbon layer at the metal-SiC interface during AIMD simulations. The overall 3D view and top view emphasize the differences in the trajectories of carbon atoms at the interface depending on the metal elements. The computed changes in the m-membered rings centered on carbon atoms and coordination numbers clearly reveal that only Ni metal stabilizes a graphene-like structure at the interface, while this stabilization effect is not evident for Fe and Ru metals.

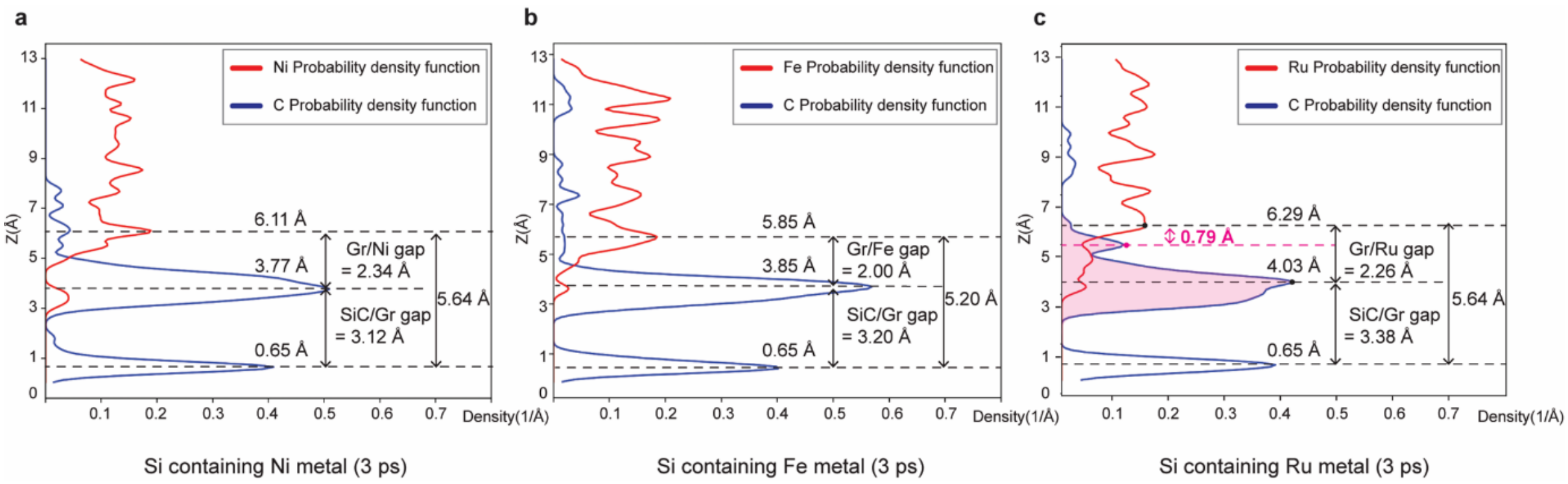

**Supplementary Fig. 6 | Study on the van der Waals gap between the metal and carbon layers using the probability-density function.** The width of the van der Waals gap observed in (a) the Si-dissolved Ni model, (b) the Si-dissolved Fe model, and (c) the Si-dissolved Ru model. The Ni-containing model maintained a consistent van der Waals gap between the metal and the topmost carbon layer, whereas the Fe- and Ru-containing models exhibited a collapse of the gap.

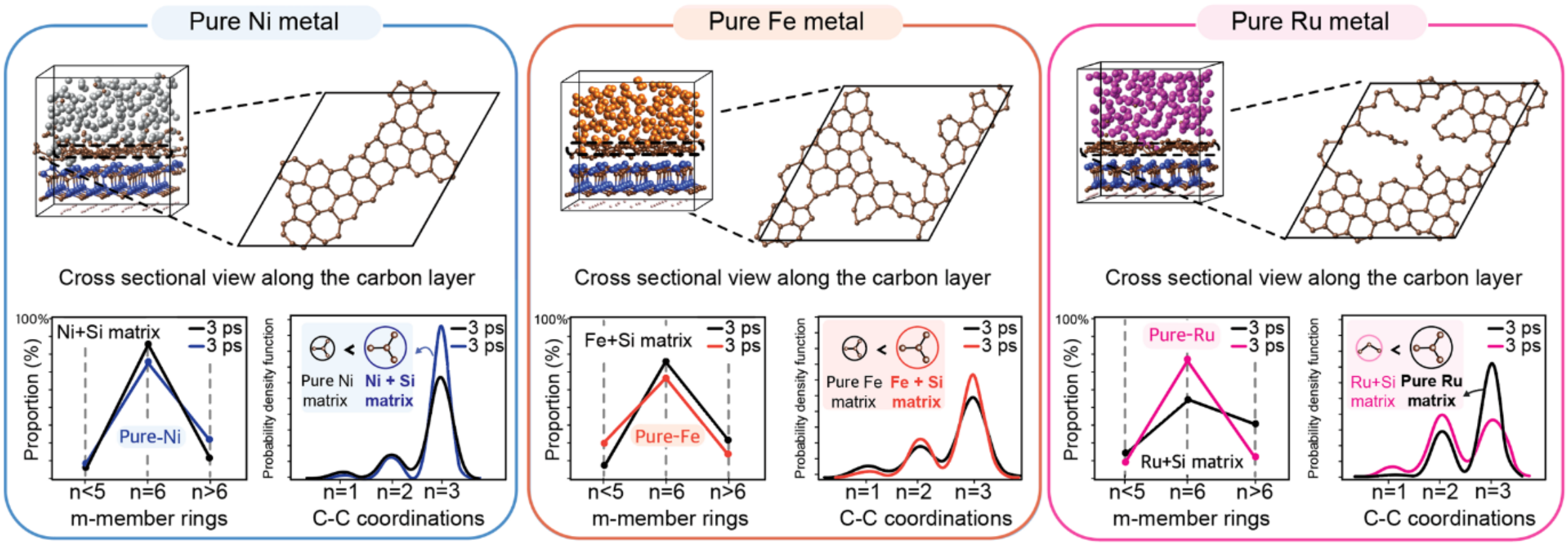

**Supplementary Fig. 7 | AIMD simulations of metal layers without Si alloying.** For pure Ni and Fe models, the number of six-fold rings and three-fold-coordinated carbon atoms was consistently lower than in their Si-dissolved counterpart, underscoring the critical role of dissolved silicon in enhancing the structural transition toward graphene-like configurations. In contrast, graphene formation in the Ru matrix occurs only when silicon is not dissolved in the Ru bulk. Among the models studied, only the Si-dissolved Ni model provided an environment where carbon atoms could readily reorganize, enabling the formation of stable, two-dimensional graphene configurations.

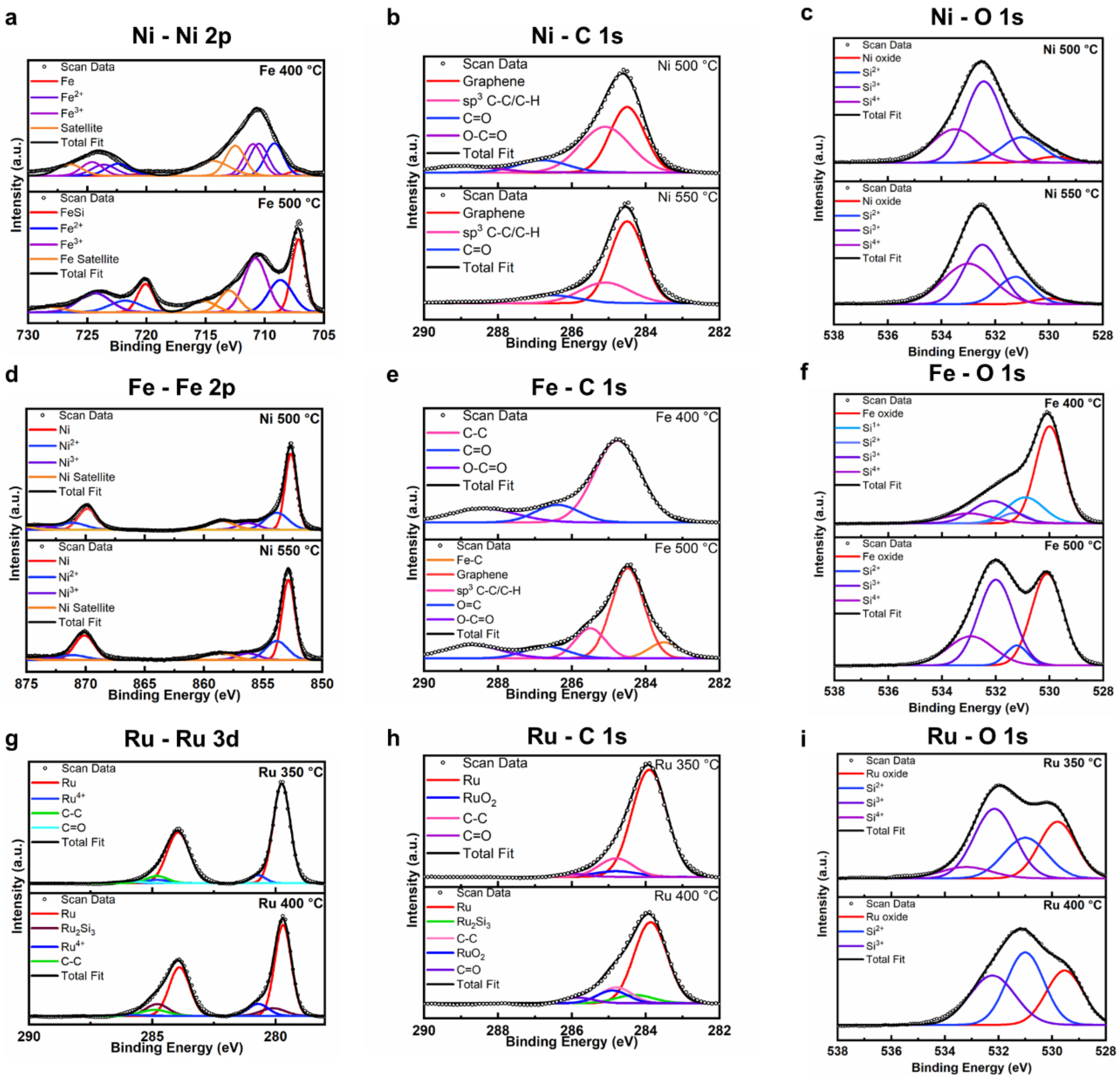

**Supplementary Fig. 8 | X-ray photoelectron spectra of metal/SiC after annealing at various temperatures. a–c,** Ni/SiC: Ni 2p, C 1s, and O 1s spectra after annealing at 500 °C and 550 °C. **d–f**, Fe/SiC: Fe 2p, C 1s, and O 1s spectra after annealing at 400 °C and 500 °C. **g–i**, Ru/SiC: Ru 3d, C 1s, and O 1s spectra after annealing at 350 °C and 400 °C. All spectra were calibrated using the C-C bond (284.8 eV) or C=C bond (284.6 eV) and cross-checked with Raman spectra (see Supplementary Fig. 9) and XRD (see Supplementary Fig. 10) results. Additional calibration was performed using the O 1s peak, which reflects oxidation induced during and after annealing. The thickness of the deposited metal layer exceeded the detection limit depth of XPS (typically around 10 nm) indicating that the detected Si and C atoms originated from the SiC.

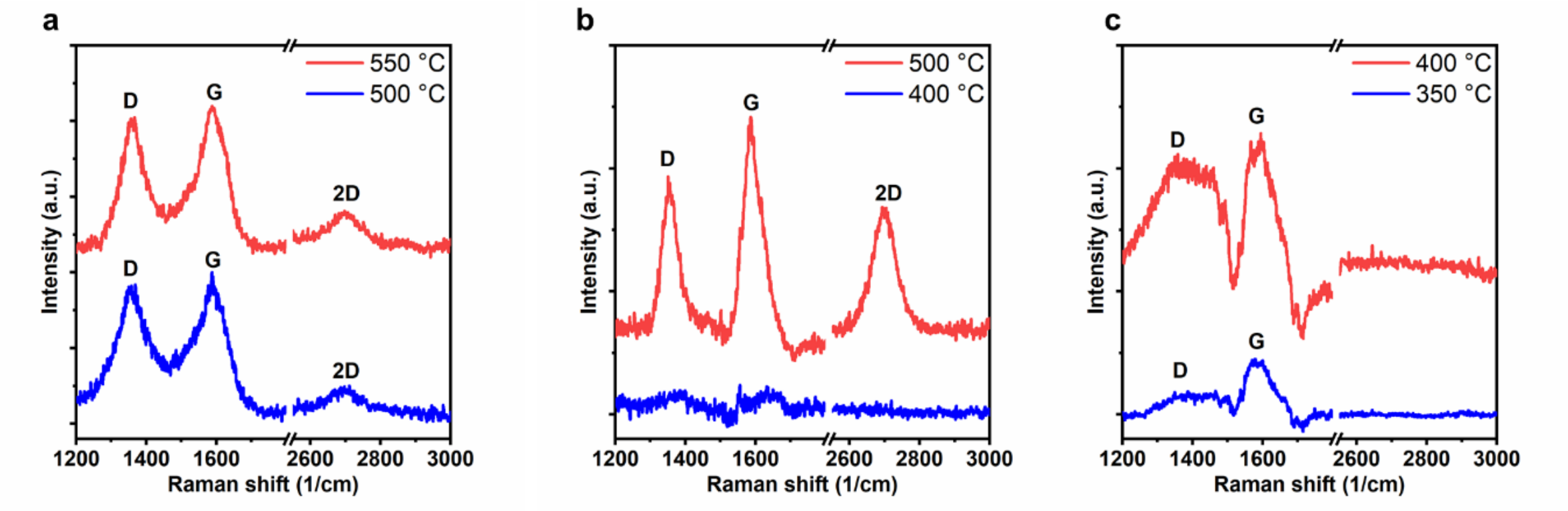

**Supplementary Fig. 9 | Raman spectra after annealing at corresponding temperatures. a,** Comparison of the Ni/SiC Raman spectra at different annealing temperatures. **b**, Comparison of the Fe/SiC Raman spectra at different annealing temperatures. **c**, Comparison of the Ru/SiC Raman spectra at different annealing temperatures. At 500 °C for Ni, 500 °C for Fe, and 400 °C for Ru, liberated carbon diffused outward and formed graphitic carbon layer on the metal surface, consistent with XPS results.

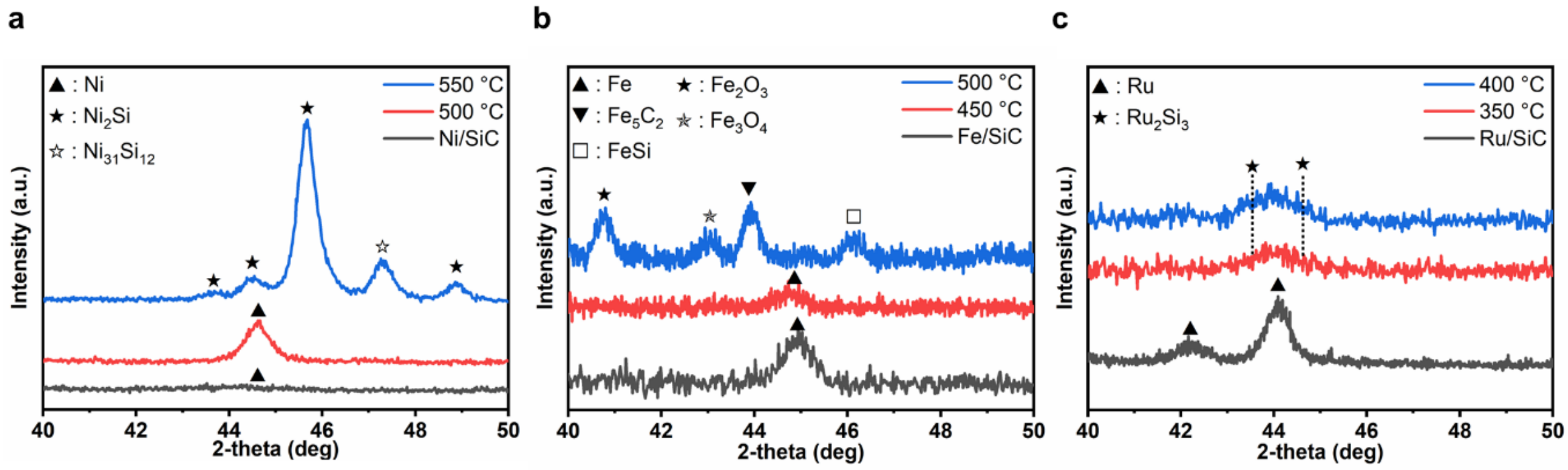

**Supplementary Fig. 10 | X-ray diffraction patterns after annealing at corresponding temperatures. a,** Comparison of the Ni/SiC X-ray diffraction patterns at various annealing temperatures. The reference patterns correspond to JCPDS no. 71-4655 for Ni, JCPDS no. 79-3559 for $Ni_2Si$, and JCPDS no. 17-0222 for $Ni_{31}Si_{12}$. **b**, Comparison of the Fe/SiC X-ray patterns at various annealing temperatures. The reference patterns correspond to JCPDS no. 76-6588 for Fe, JCPDS no. 79-3559 for FeSi, JCPDS no. 89-8103 for Fe2O3, JCPDS no. 71-6336 for $Fe_3O_4$, and JCPDS no. 89-8968 for $Fe_5C_2$. **c**, Comparison of the Ru/SiC X-ray diffraction patterns at various annealing temperatures. The reference patterns correspond to JCPDS no. 73-7011 for Ru, and JCPDS no. 88-0895 for $Ru_2Si_3$.

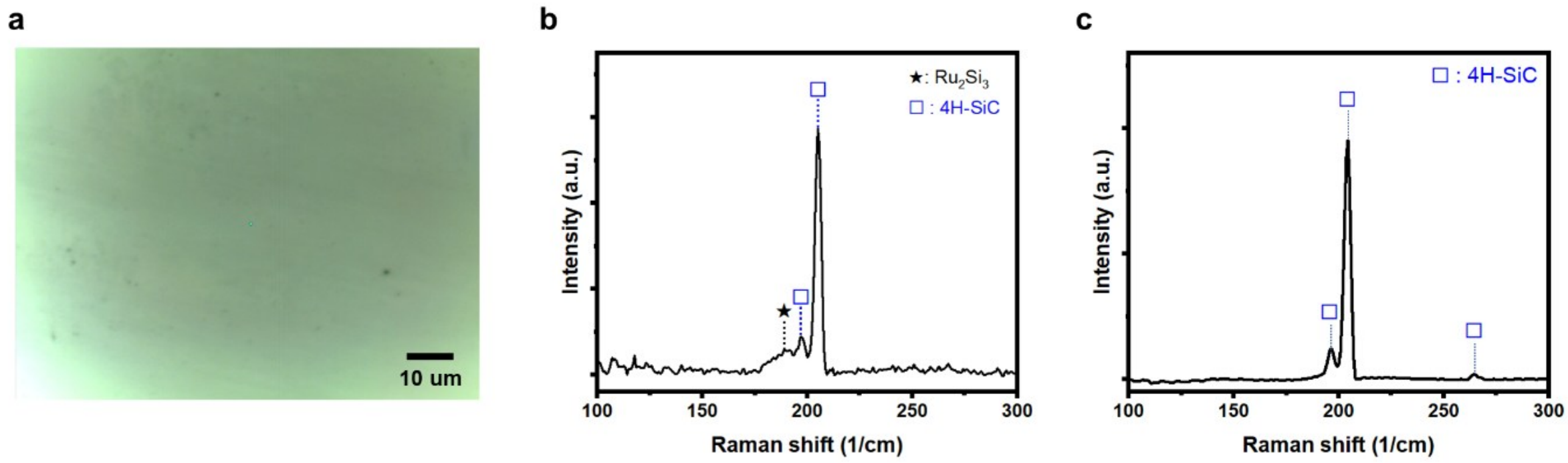

**Supplementary Fig. 11 | Optical image and Raman spectra of the Ru/SiC annealed at 350 °C. a-b,** Raman spectra with optical image reveal the presence of $Ru_2Si_3$ peak at the 203 $cm^{-1}$, consistent with XPS and XRD results. **c,** Raman spectra of bare SiC in range of 50 $cm^{-1}$ to 500 $cm^{-1}$, indicate the absence of the 203 $cm^{-1}$ peak.

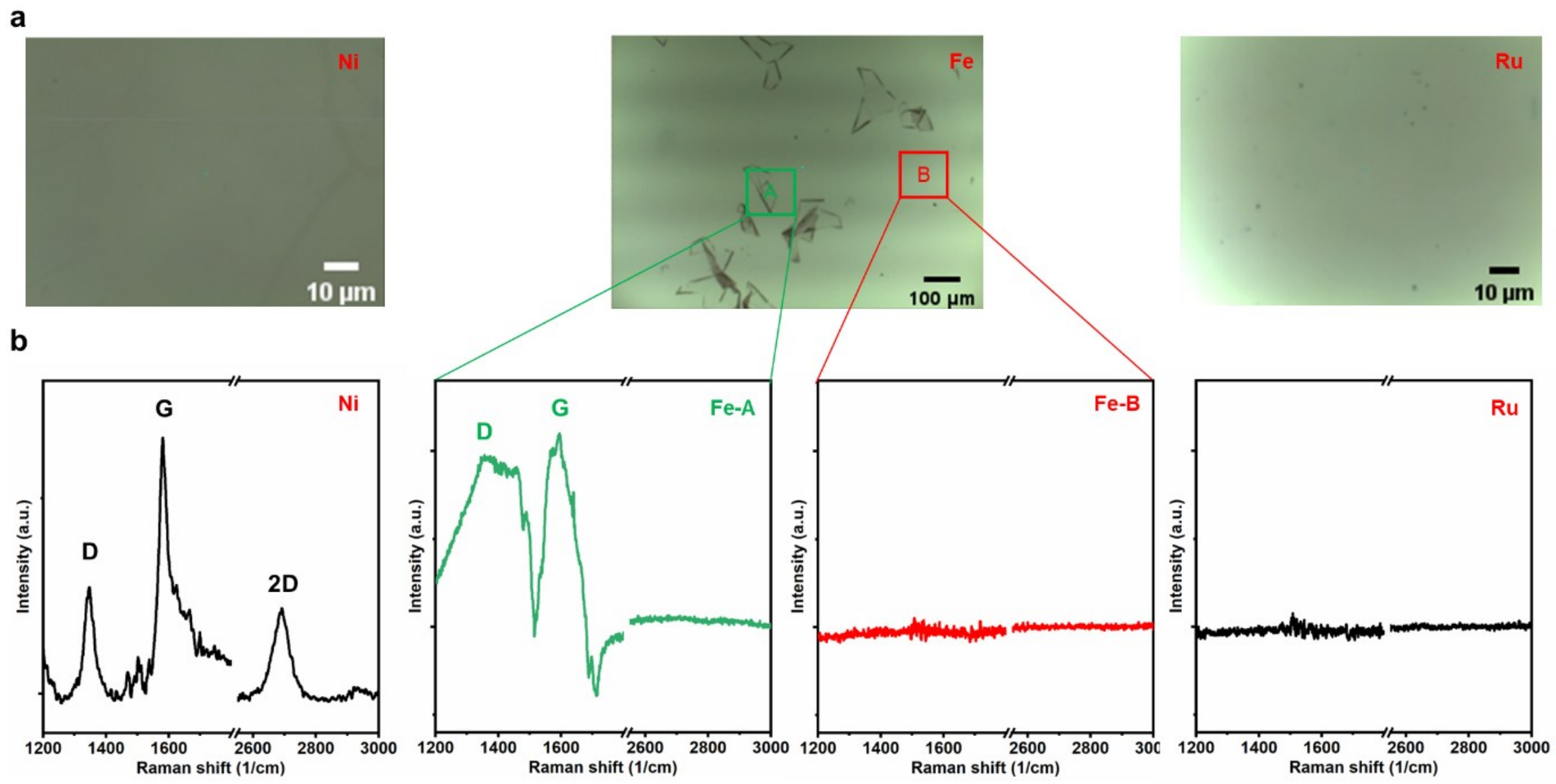

**Supplementary Fig. 12 | Micrography and Raman spectra of various metals on SiC after annealing at corresponding temperatures and etching with the appropriate etchant. a,** Micrography of Ni/SiC annealed at 500 °C and etched with $FeCl_3$, Fe/SiC annealed at 400 °C and etched with $FeCl_3$, and Ru/SiC annealed at 350 °C and etched with NaOCl. **b,** Raman spectra of each sample. The D and G bands are clearly observed in both the Ni and Fe systems. However, the absence of a discernible 2D peak and the high defect ratio ($I_D/I_G \approx 0.98$) with spatially non-uniform characteristics in the Fe system indicate that graphene formation occurs exclusively in the Ni system.

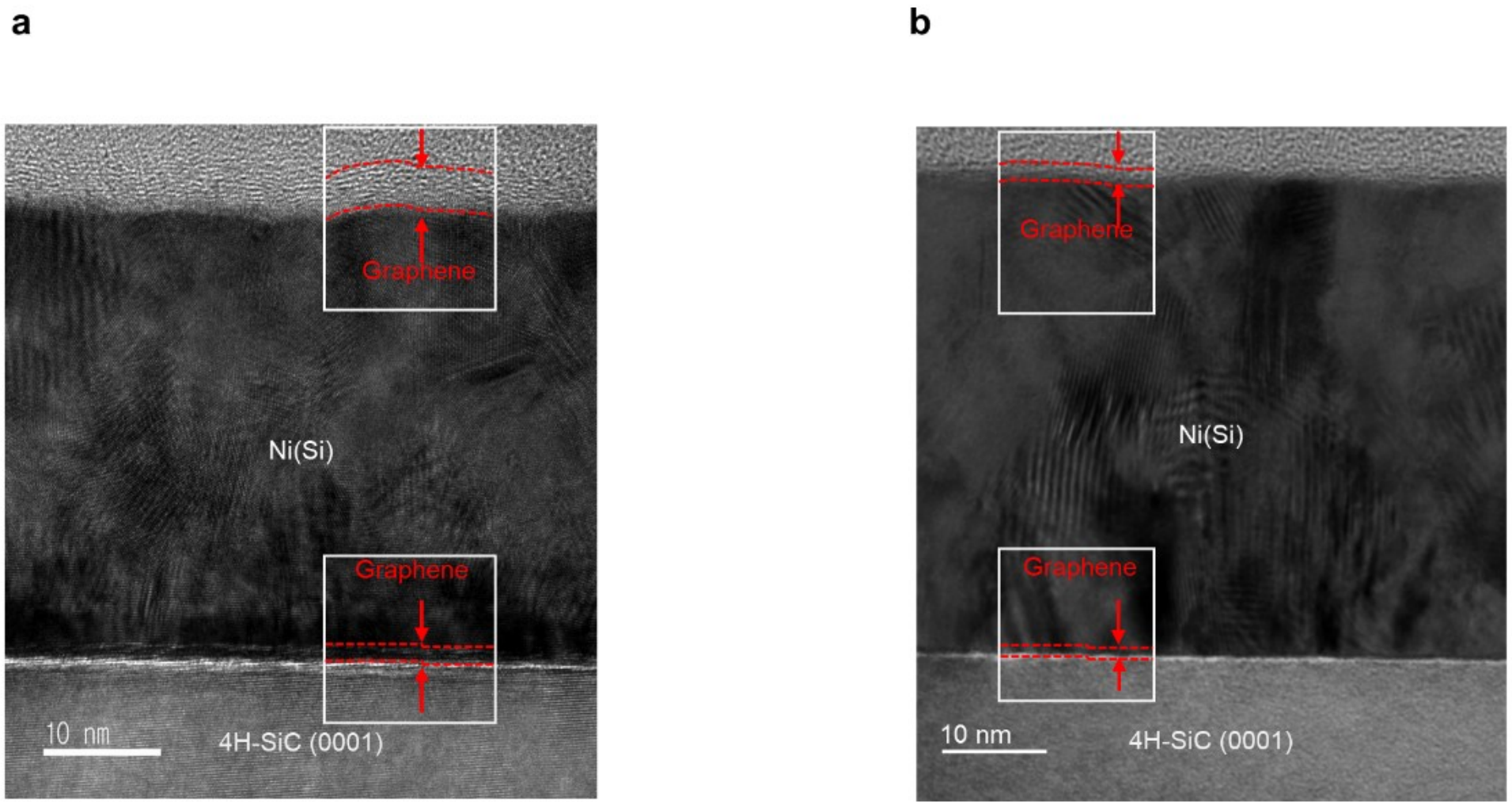

**Supplementary Fig. 13 | TEM images of Ni/SiC structure annealed at 500 °C and 320 °C respectively, measured at 10 nm scale. a,** TEM images of Ni/SiC annealed at 500 °C. **b,** TEM images of Ni/SiC annealed at 320 °C. The TEM images show the graphene layer not only on the metal surface, but also at the Ni/SiC interface. Both results suggest that graphene is slightly thicker on the Ni surface than at the Ni/SiC interface. This implies that liberated carbon atoms preferentially diffuse along the grain boundaries of Ni rather than across the Ni/SiC interface. The thickness of graphene increases on both the Ni surface and at the Ni/SiC interface as the annealing temperature rises, indicating that the MAG process is primarily driven by temperature.

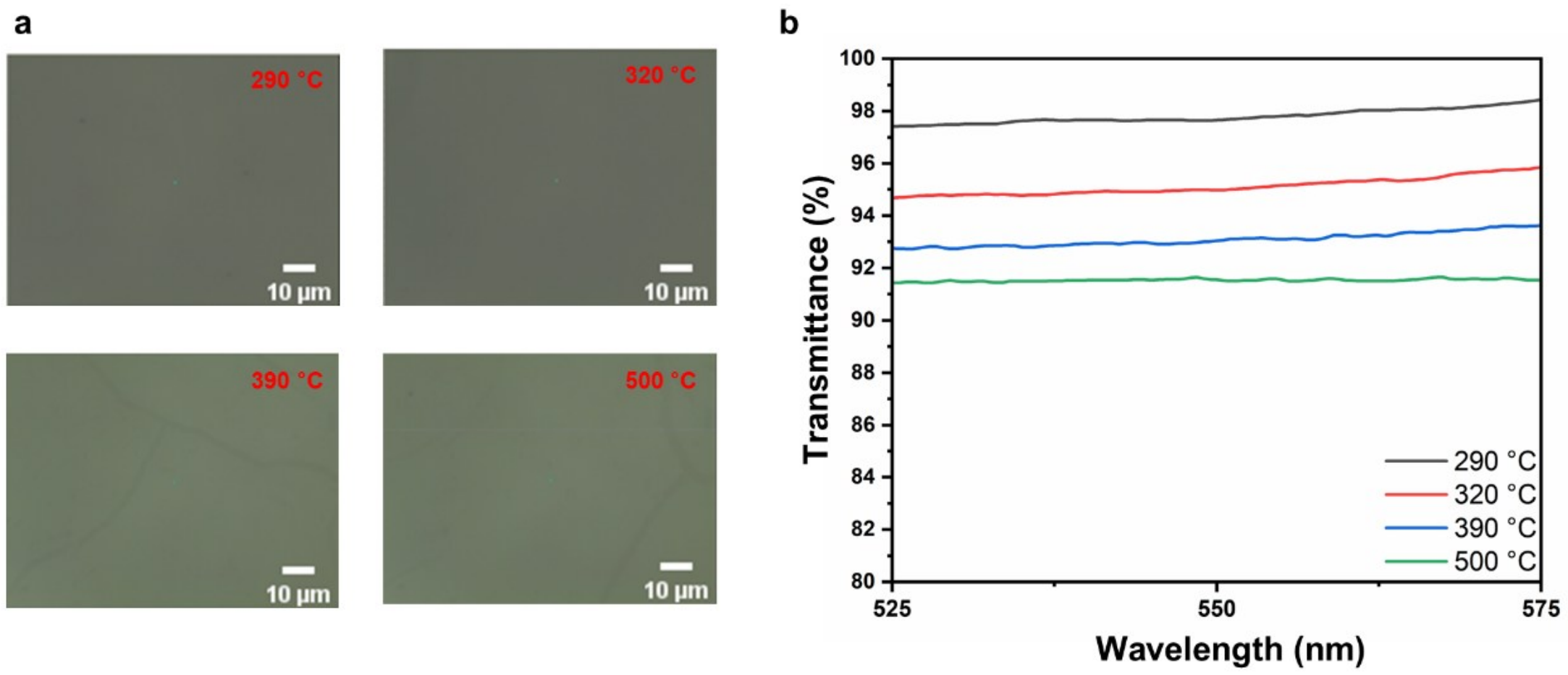

**Supplementary Fig. 14 | Optical images and UV-visible spectra after annealing at the corresponding temperatures and etching with $FeCl_3$. a,** Optical images after annealing at corresponding temperatures and etching, confirming the absence of residuals and the formation of a macroscopic graphene film. **b,** UV-visible spectra showing that the MAG process is driven by temperature.

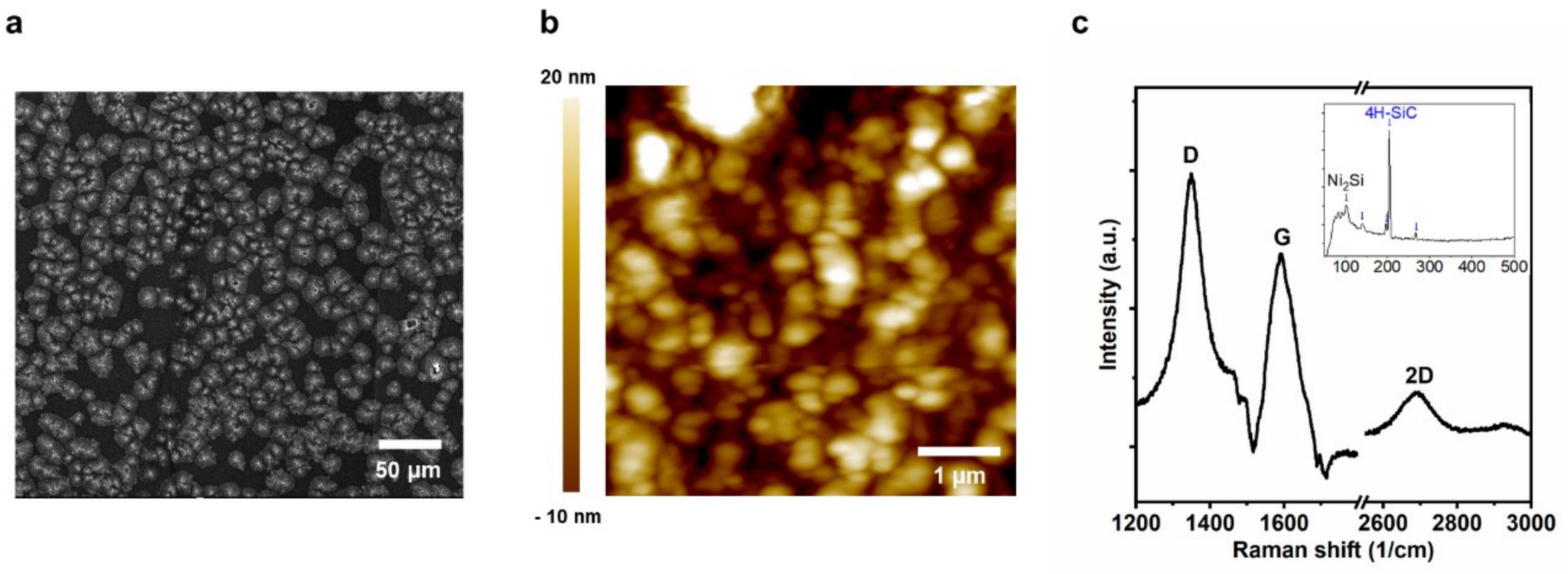

**Supplementary Fig. 15 | SEM, AFM images, and Raman spectra after annealing above the silicide formation temperature, and etching the Ni layer with $FeCl_3$. a-b,** SEM, and AFM images reveal inhomogeneous reactions, confirming the formation of a non-uniform clustered film. **c,** Raman spectra indicate that the clustered films comprise graphene on a silicide layer.

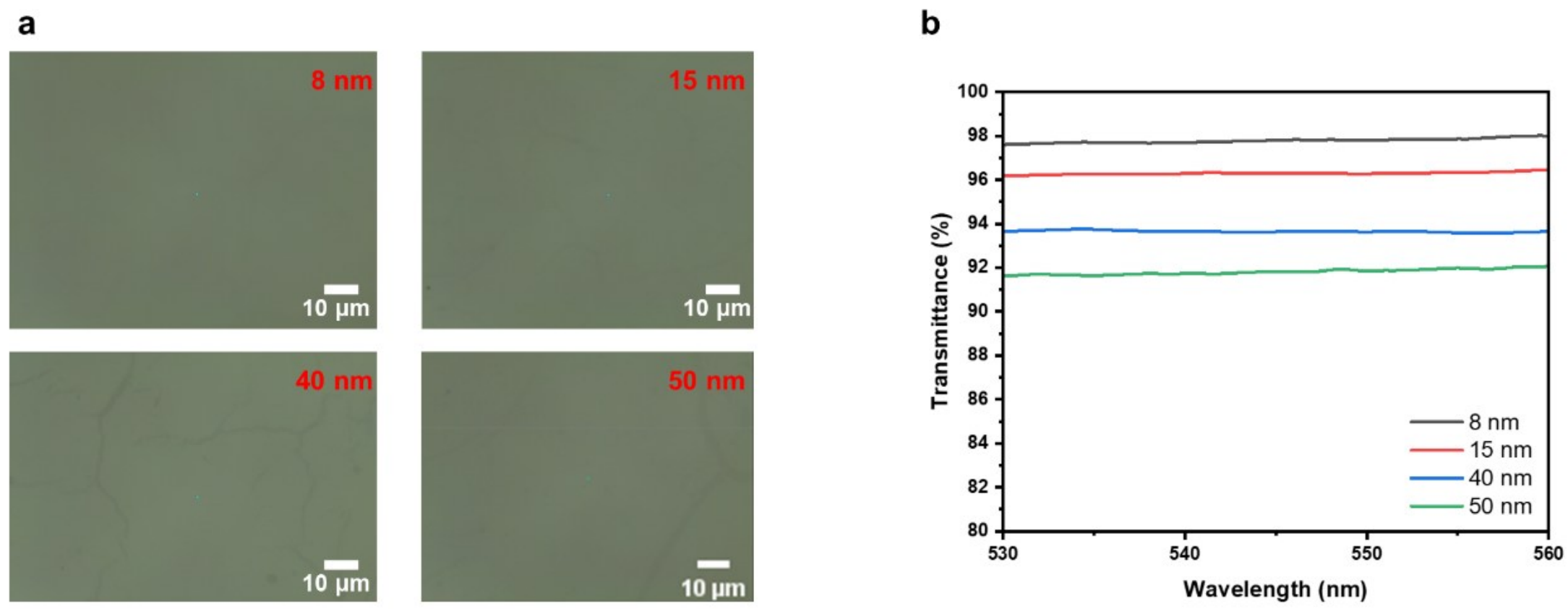

**Supplementary Fig. 16 | Optical images and UV-visible spectra for various metal thickness used in MAG process. a,** Optical images of samples with different metal thicknesses. **b,** UV-visible spectra indicating that graphene thickness decreases with reduced metal thickness.

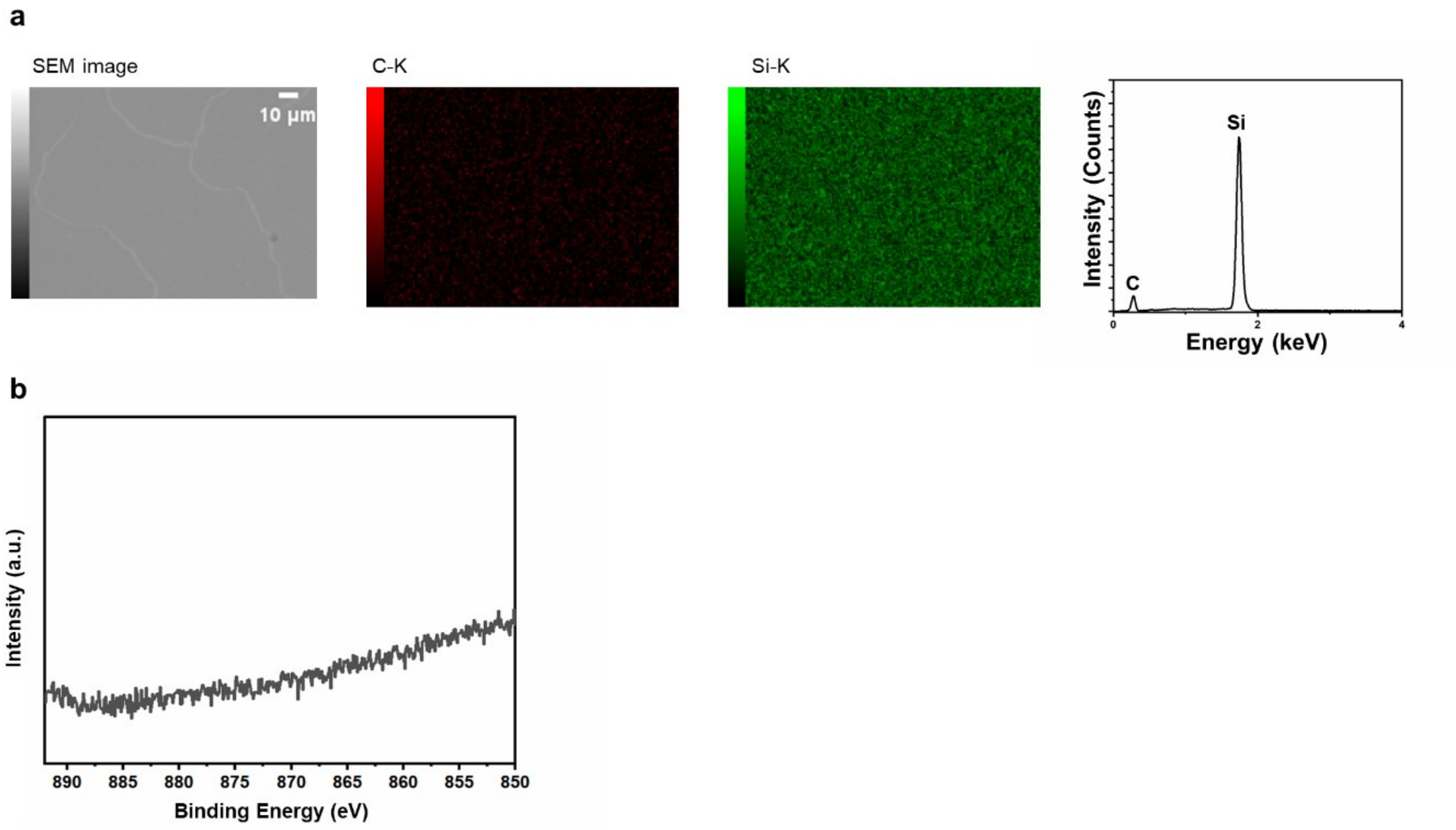

**Supplementary Fig. 17 | Energy-dispersive X-ray (EDS) spectra and Ni 2p XPS after etching the unreacted Ni layer with $FeCl_3$. a,** EDS scan shows no Ni atoms remaining on the surface. **b,** XPS Ni 2p spectra confirm the absence of detectable Ni. All results indicate the complete removal of Ni.

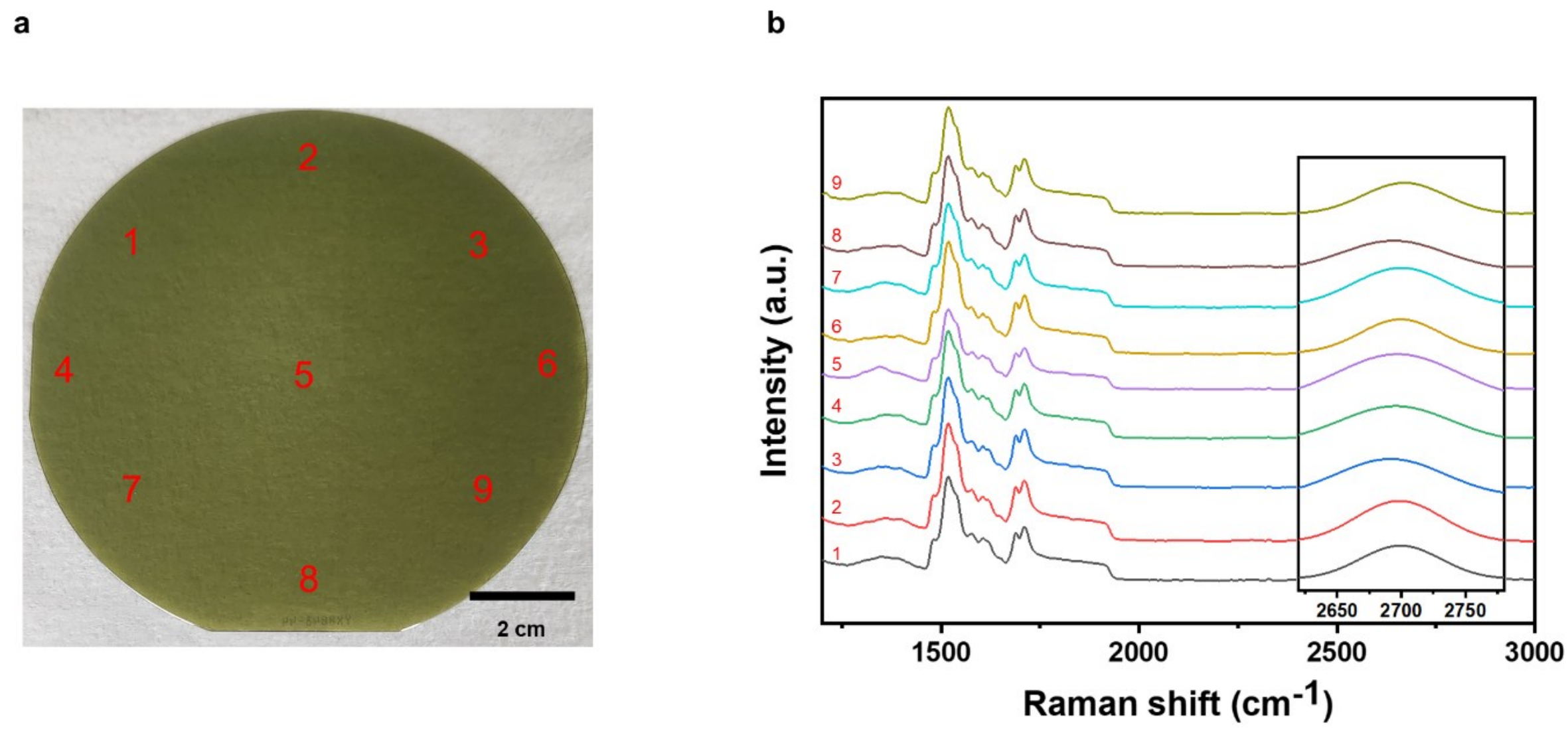

**Supplementary Fig. 18 | Raman spectra of a graphitized 4-inch wafer.** The wafer was divided into nine areas, and Raman spectra from all regions consistently show the characteristics D, G, and 2D peaks of graphene.

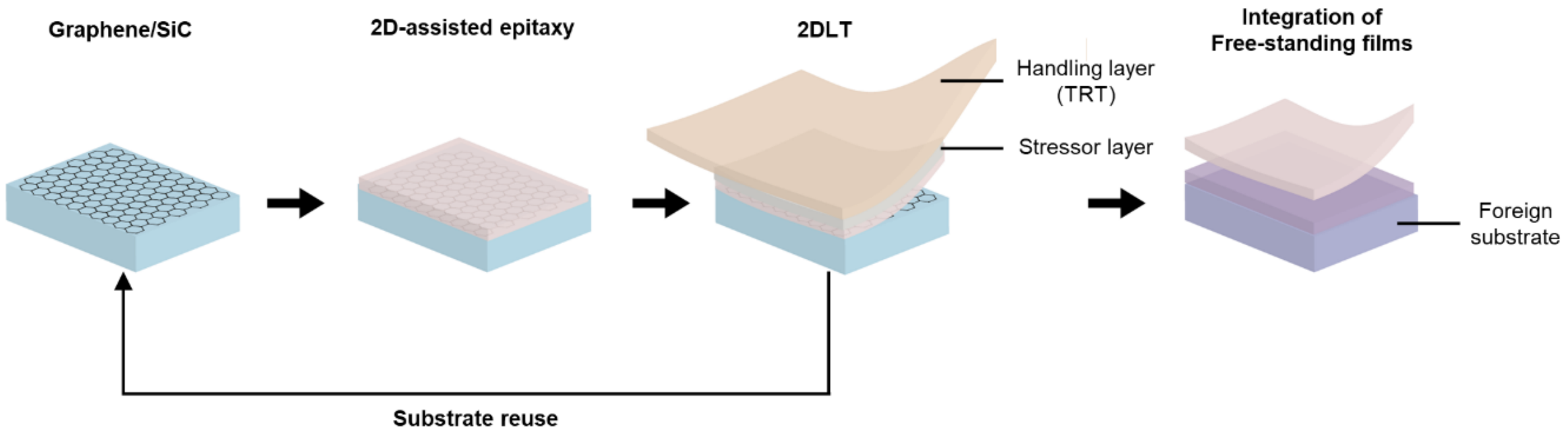

**Supplementary Fig. 19 | Schematic illustration of 2D-assisted epitaxy process.** The process begins with the graphitization of the SiC to synthesize graphene. This graphitized template is then prepared for 2D-assisted epitaxy, where a wide-bandgap material is grown using a growth chamber, such as MOCVD or MBE, under optimized conditions. After the growth, a stressor layer, typically Ni, is deposited via a sputtering or E-beam evaporator. The thickness and deposition conditions of the Ni layer must be carefully controlled to enable precise exfoliation of the membrane at the van der Waals gap between the membrane and the graphitized SiC interface. Additionally, an adhesion layer may be employed to prevent peeling of the stressor layer. A TRT is then applied as a handling layer, and exfoliation is carried out by smoothly lifting the handling layer. The exfoliated freestanding membrane is subsequently transferred to a foreign substrate. Finally, the TRT is removed by heating the membrane-attached-foreign-sample above 110 °C.

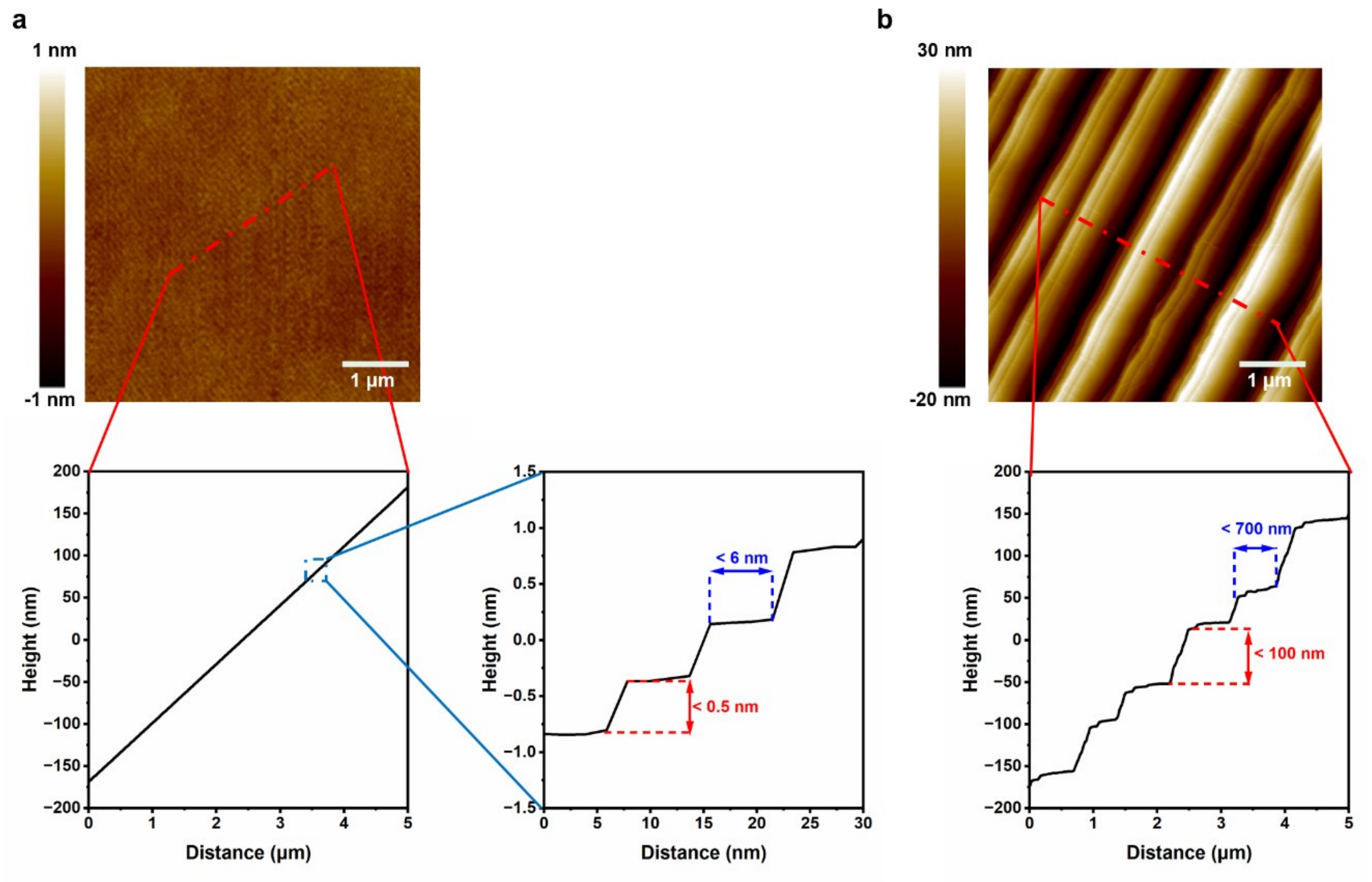

**Supplementary Fig. 20 | Comparison of AFM images and line profiles of high-temperature graphitized SiC and MAG-treated SiC. a,** AFM images and corresponding line profiles (along the red line) reveal terrace widths below 6 nm and step heights of approximately 0.5 nm. **b,** AFM images and corresponding line profiles (along the red line) show terrace widths on the micrometer scale and step heights around 700 nm.

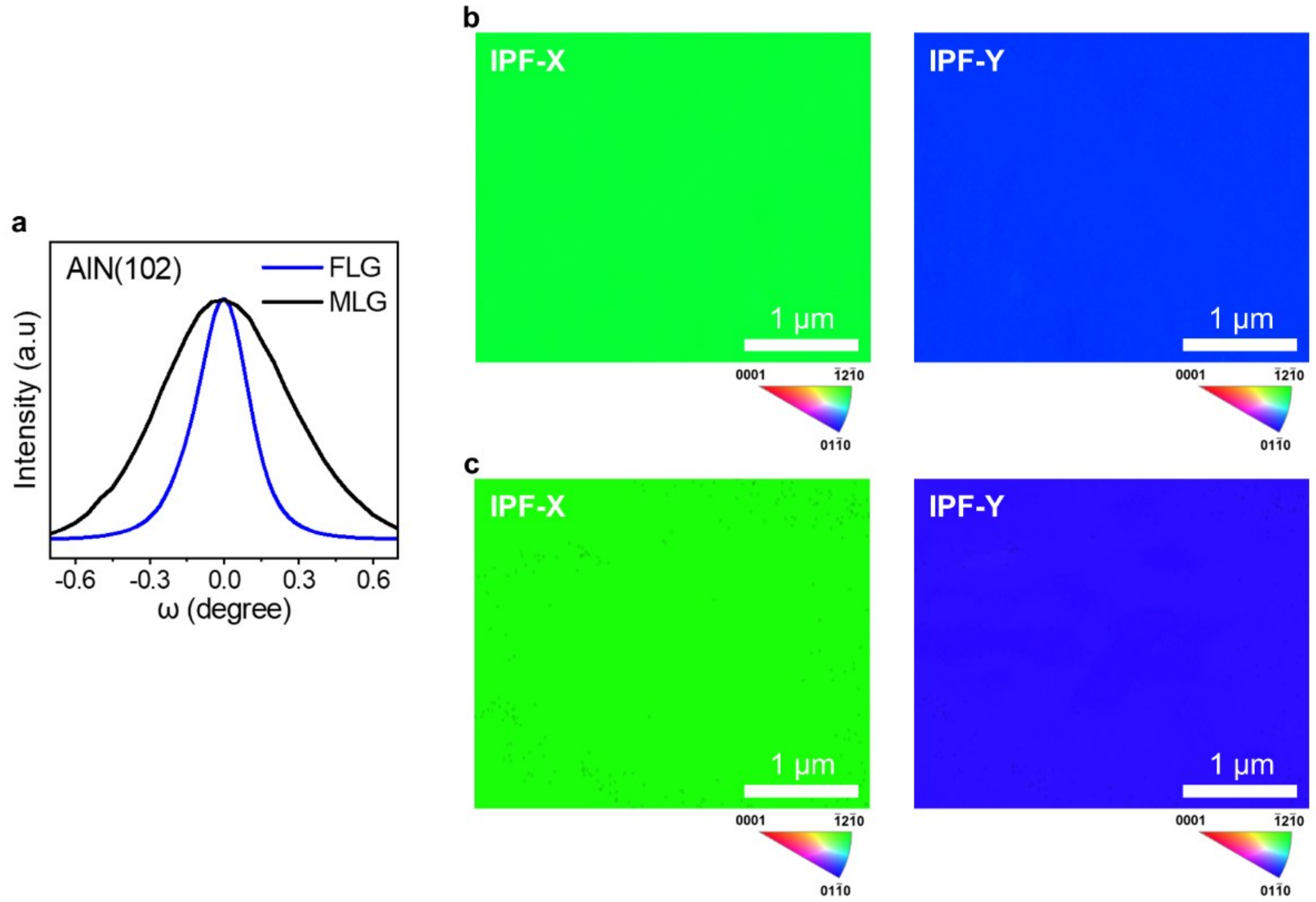

**Supplementary Fig. 21 | HR-XRD rocking curve and In-plane EBSD of AlN grown on MAG-treated FLG and MLG on SiC. a,** In-plane XRD rocking curve of the (102) plane, showing a FWHM of 833 arcsec for AlN on FLG/SiC and 2136 arcsec for AlN on MLG/SiC, comparable to the (002) plane, which exhibits FWHM values of 759 arcsec and 1744 arcsec, respectively. **b,** In-plane EBSD maps of AlN grown on FLG/SiC. **c** In-plane EBSD maps of AlN grown on MLG/SiC. The EBSD patterns indicate the single-crystalline nature of the AlN film on both FLG/SiC and MLG/SiC templates.

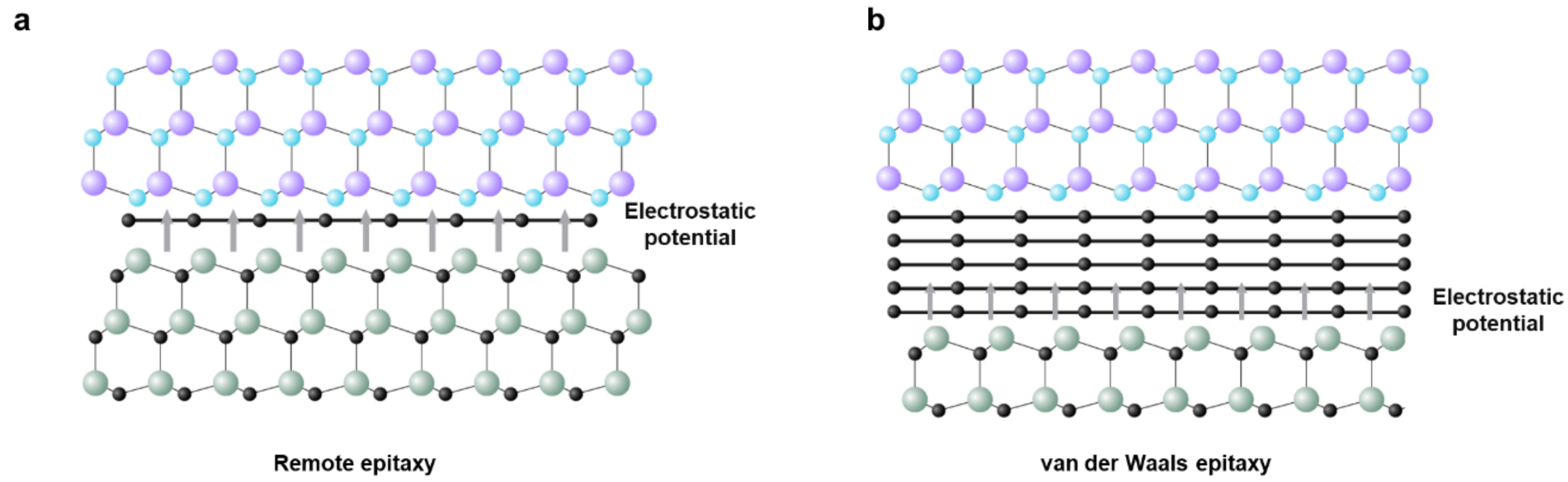

**Supplementary Fig. 22 | Schematic illustration of 2D-assisted epitaxy mechanisms.** 2D-assisted epitaxy encompasses techniques like remote epitaxy and vdWE, which enable high-quality material growth on 2D materials. However, these techniques differ significantly in their requirements for preparing the graphitized sample. Remote epitaxy requires a minimum graphene layer that is robust enough to withstand the harsh growth environment while allowing the lattice information of the underlying substrate to guide the growth of the crystalline film. In contrast, vdWE does not rely on the lattice information of the substrate. Instead, it demands multi-layers of graphene to completely screen the substrate's electrostatic potential, which enhances the quality of the grown membrane by eliminating substrate interference.

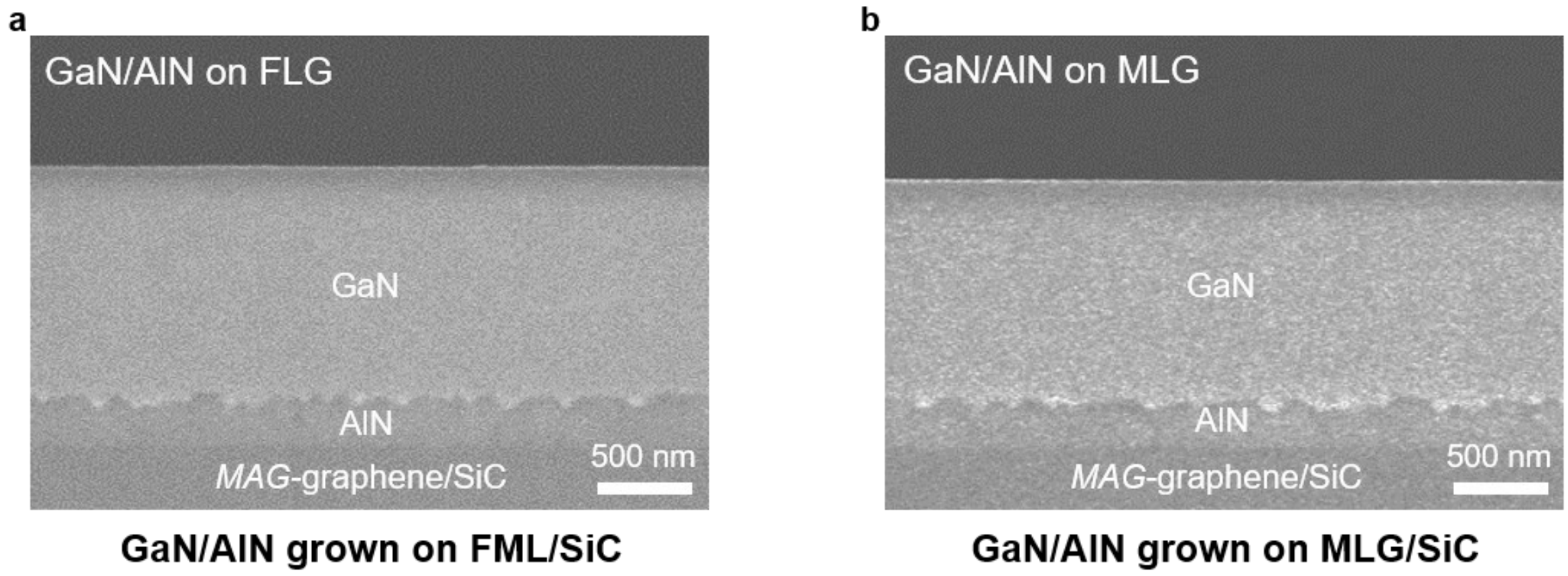

**Supplementary Fig. 23 | GaN/AlN heterostructure film grown on MAG-treated FLG/SiC and MLG/SiC. a,** SEM Cross-sectional image of GaN/AlN grown on FLG/SiC. **b,** SEM Cross-sectional image of GaN/AlN grown on MLG/SiC. Both GaN/AlN heterostructure films were epitaxially grown under identical MOCVD conditions.

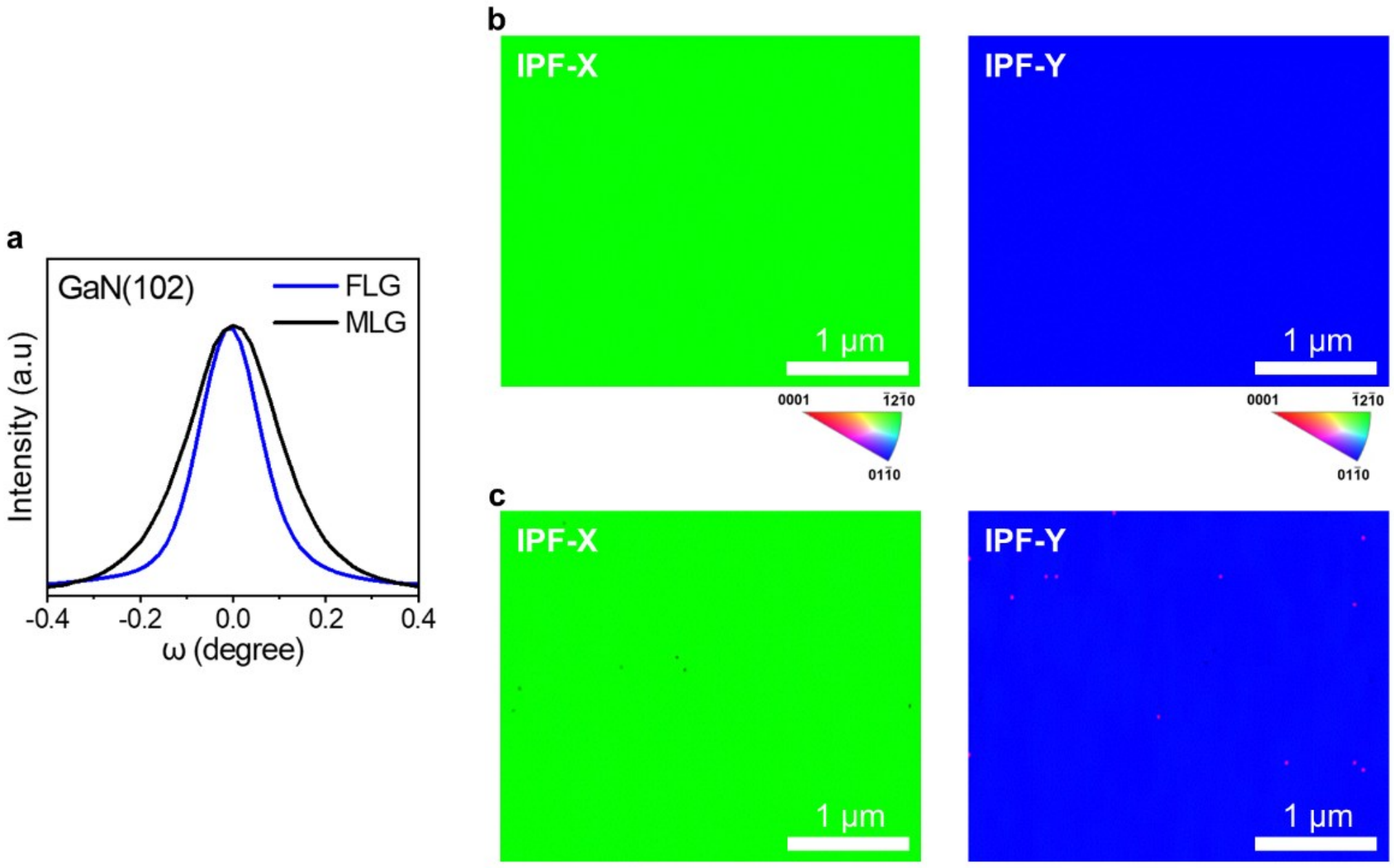

**Supplementary Fig. 24 | HR-XRD rocking curve and In-plane EBSD of GaN/AlN grown on MAG-treated FLG/SiC and MLG/SiC. a,** In-plane HR-XRD rocking curve of the (102) plane, showing a FWHM of 509 arcsec for GaN on AlN/FLG/SiC and 760 arcsec for GaN on AlN/MLG/SiC, comparable to the (002) plane, which exhibits FWHM values of 270 arcsec and 444 arcsec, respectively. **b,** In-plane EBSD maps of GaN grown on AlN/FLG/SiC. **c** In-plane EBSD maps of GaN grown on AlN/MLG/SiC. The EBSD patterns indicate the single-crystalline nature of the GaN film on both AlN/FLG/SiC and AlN/MLG/SiC templates.

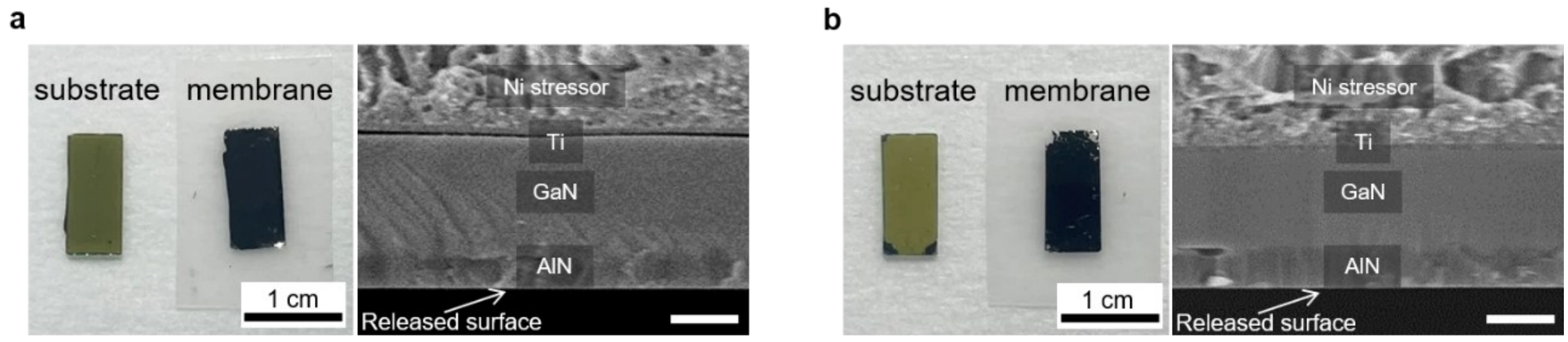

**Supplementary Fig. 25 | Camera and SEM cross-sectional images of exfoliated GaN/AlN membrane grown on MAG-treated SiC templates. a,** Exfoliation results of GaN/AlN grown on FLG/SiC. **b,** Exfoliation results of GaN/AlN grown on MLG/SiC. The camera images confirm the successful exfoliation of the GaN/AlN membrane from MAG-treated SiC templates, while the SEM cross-sectional images reveal macroscopically smooth surfaces of the exfoliated films, validating the effectiveness of the 2DLT process. The scale bar in the SEM images is 500 nm.

**Supplementary Table. 1 | Comparative analysis of AlN crystalline quality on MAG-treated SiC versus 2D-assisted epitaxial substrates.** The table highlights the significantly improved crystalline quality of free-standing AlN grown on MAG-treated SiC, compared to other 2D-material-assisted epitaxial substrates. Notably, our graphene layer does not induce defects and can be easily transferred to other substrates, which is highly advantageous for substrate reuse and enables the use of the membrane as a freestanding device. LT refers to low-temperature growth; TDD refers to threading dislocation density.

| Material | 2D-material | 2D material growth method | Substrate | AlN Growth | (002) / (102) FWHM (arcsec) | TDD ($cm^{-2}$) | Exfoliation | Ref |
|---|---|---|---|---|---|---|---|---|
| AlN | Multi-layer Graphene | High temp graphitization | SiC | HVPE (vdWE) | 3600 / - | - | ○ | [5] |
| | Defect induced Graphene | CVD (1055 °C, 3-5 h) | $Al_2O_3$ | MOCVD (qvdWE) | 315.5 / 518 | $2.7 \times 10^9$ | × | [6] |
| | Defect induced Graphene | High temp graphitization (> 1600 °C) | SiC | MOCVD (qvdWE) | 148 / 79 | $6.7 \times 10^8$ | × | [7] |
| | *Multi-layer Graphene* | *MAG (450 °C)* | *SiC* | MOCVD (vdWE) | *1774 / 2136* | $3.1 \times 10^{10}$ | *O* | *This work* |
| | *Few-layer Graphene* | *MAG (320 °C)* | *SiC* | MOCVD (Remote Epitaxy) | *759 / 833* | $3.8 \times 10^9$ | *O* | *This work* |

**Supplementary Table. 2 | Comparison of the crystalline quality of GaN/AlN heterostructures on MAG-treated SiC versus other substrates.** The results suggest that our MAG-treated SiC offers a desirable platform for GaN materials. LT refers to low-temperature growth.

| Material | Buffer layer | Substrate | Growth Method | (002) FWHM (arcsec) | Threading Dislocation Density ($cm^{-2}$) | Reference |
|---|---|---|---|---|---|---|
| GaN | AlN | Si (111) | MOCVD | 380 | $3 \times 10^9$ | [8] |
| | LT-AlN | $Al_2O_3$ | MOCVD | 380 | $6 \times 10^8$ | [9] |
| | LT-GaN | $Al_2O_3$ | MOCVD | 220 | $9 \times 10^8$ | [9] |
| | AlN / h-BN (3nm) | $Al_2O_3$ | MOCVD | 576 | $1.6 \times 10^{10}$ | [10] |
| | AlN / h-BN | $Al_2O_3$ | MOCVD | 396 | $3 \times 10^8$ | [11] |
| | AlN | SiC | MOCVD | 200 | $2 \times 10^9$ | [12] |
| | Graphene buffer layer | SiC | MBE | 141 | $2.1 \times 10^7$ | [13] |
| | 1 ML Graphene | SiC | MBE | 328 | $6.3 \times 10^8$ | [13] |
| | LT-GaN (150 nm) / Graphene buffer layer | SiC | MOCVD | 544 | $3.0 \times 10^9$ | [14] |
| | 1 ML Graphene | SiC | MOCVD | 222 | $1 \times 10^9$ | [15] |
| | AlN / Multi-layer Graphene | SiC | MOCVD | 1260 | $1.4 \times 10^{10}$ | [16] |
| | *AlN / MLG* | *SiC* | MOCVD | *270* | $1.53 \times 10^9$ | *This work* |
| | *AlN / FLG* | *SiC* | MOCVD | *444* | $4.07 \times 10^9$ | *This work* |

## Supplementary References

[1] D. Dardzinski, M. Yu, S. Moayedpour, N. Marom, *J. Phys.: Condens. Matter* **2022**, *34*, DOI 10.1088/1361-648X/ac577b.

[2] B. P. Nash, A. Nash, *Bull. Alloy Phase Diagr.* **1987**, *8*, 6.

[3] K. Vojtčrtovský, T. Zemčík, *Czech. J. Phys. B* **1974**, *24*, 171.

[4] L. Perring, F. Bussy, J. C. Gachon, P. Feschotte, *J. Alloys Compd.* **1999**, *284*, 198–205.

[5] Y. Xu, B. Cao, Z. Li, D. Cai, Y. Zhang, G. Ren, J. Wang, L. Shi, C. Wang, K. Xu, *ACS Appl. Mater. Interfaces* **2017**, *9*, 44001.

[6] Z. Chen, Z. Liu, T. Wei, S. Yang, Z. Dou, Y. Wang, H. Ci, H. Chang, Y. Qi, J. Yan, J. Wang, Y. Zhang, P. Gao, J. Li, Z. Liu, *Adv. Mater.* **2019**, *31*, 1807345.

[7] Y. Wang, S. Yang, H. Chang, W. Li, X. Chen, R. Hou, J. Yan, X. Yi, J. Wang, T. Wei, *Appl. Surf. Sci.* **2020**, *520*, 146358.

[8] E. Arslan, M. K. Ozturk, A. Teke, S. Ozcelik, E. Ozbay, *J. Phys. D: Appl. Phys.* **2008**, *41*, 155317.

[9] C. Bayram, J. L. Pau, R. McClintock, M. Razeghi, *J. Appl. Phys.* **2008**, *104*, 083512.

[10] Y. Kobayashi, K. Kumakura, T. Akasaka, T. Makimoto, *Nature* **2012**, *484*, 223–227.

[11] N. R. Glavin, K. D. Chabak, E. R. Heller, E. A. Moore, T. A. Prusnick, B. Maruyama, D. E. Walker, D. L. Dorsey, Q. Paduano, M. Snure, *Adv. Mater.* **2017**, *29*, 1701838.

[12] Z. J. Reitmeier, S. Einfeldt, R. F. Davis, X. Zhang, X. Fang, S. Mahajan, *Acta Mater.* **2010**, *58*, 2165–2175.

[13] K. Qiao, Y. Liu, C. Kim, R. J. Molnar, T. Osadchy, W. Li, X. Sun, H. Li, R. L. Myers-Ward, D. Lee, S. Subramanian, H. Kim, K. Lu, J. A. Robinson, W. Kong, J. Kim, *Nano Lett.* **2021**, *21*, 4013–4020.

[14] S. Lee, J. Kim, B. I. Park, H. I. Kim, C. Lim, E. Lee, J. Y. Yang, J. Choi, Y. J. Hong, C. S. Chang, H. S. Kum, J. Kim, K. Lee, H. Kim, G. C. Yi, *Appl. Phys. Lett.* **2024**, *125*, 252102.

[15] J. Kim, C. Bayram, H. Park, C. W. Cheng, C. Dimitrakopoulos, J. A. Ott, K. B. Reuter, S. W. Bedell, D. K. Sadana, *Nat. Commun.* **2014**, *5*, 4836.

[16] Y. Yu, T. Wang, X. Chen, L. Zhang, Y. Wang, Y. Niu, J. Yu, H. Ma, X. Li, F. Liu, G. Deng, Z. Shi, B. Zhang, X. Wang, Y. Zhang, *Light Sci. Appl.* **2021**, *10*, 60.

**Other Supplementary Materials for this manuscript include:**

**Supplementary Video 1. AIMD simulation of the Ni/graphitic carbon/SiC system.**

**Supplementary Video 2. AIMD simulation of the Fe/graphitic carbon/SiC system.**

**Supplementary Video 3. AIMD simulation of the Ru/graphitic carbon/SiC system.**